\documentclass[prd,aps,amsmath,amssymb,showpacs]{revtex4}
\usepackage{graphicx}
\usepackage{epsfig,rotate,latexsym}
\usepackage{hyperref}
\begin{document}
\title{Simulation of Z(3) walls and string production via bubble 
nucleation in a quark-hadron transition}
\author{Uma Shankar Gupta$^1$}
\email {guptausg@gmail.com}
\author{Ranjita K. Mohapatra$^2$}
\email {ranjita@iopb.res.in}
\author{Ajit M. Srivastava$^2$}
\email{ajit@iopb.res.in}
\author{Vivek K. Tiwari$^1$}
\email{vivek_krt@hotmail.com}
\affiliation{$^1$Physics Department, Allahabad University, 
Allahabad 211002, India \\
$^2$Institute of Physics, Sachivalaya Marg, Bhubaneswar 751005, India}
%
%

\begin{abstract}

 We study the dynamics of confinement-deconfinement (C-D) phase transition 
in the context of relativistic heavy-ion collisions within the framework
of effective models 
for the Polyakov loop order parameter. We study the formation of $Z(3)$ 
walls and associated strings in the initial transition from the 
confining (hadronic) phase to the deconfining (QGP) phase via the
so called Kibble mechanism. Essential physics of the Kibble mechanism
is contained in a sort of domain structure arising after any phase
transition which represents random variation of the order parameter 
at distances beyond the typical correlation length. We implement this
domain structure by using the Polyakov loop effective model 
with a first order phase transition and confine ourselves with 
temperature/time ranges so that the first order C-D transition
proceeds via bubble nucleation, leading to a well defined domain
structure. The formation of $Z(3)$ walls and associated strings 
results from the coalescence of QGP bubbles expanding in the confining 
background. We investigate the evolution of the $Z(3)$ wall and 
string network. We also calculate the energy density fluctuations 
associated with $Z(3)$ wall network and strings which decay away
after the temperature drops below the quark-hadron transition 
temperature during the expansion of QGP. We  discuss
evolution of these quantities with changing temperature via Bjorken's
hydrodynamical model and discuss possible experimental
signatures resulting from the presence of $Z(3)$ wall network and
associate strings.

\end{abstract}

\maketitle

 \section{INTRODUCTION}

 Search for the quark-gluon plasma (QGP) at relativistic heavy-ion
collision experiments (RHICE) has reached a very exciting stage with the 
ongoing experiments at RHIC and with the upcoming heavy-ion experiments
at the Large Hadron Collider.
No one doubts that the QGP phase has already been created at RHIC
but conclusive evidence for the same is still lacking. Many
signatures have been proposed for the detection of the QGP phase
\cite{heinz} and these have been thoroughly investigated both 
theoretically and experimentally. Along with continued investigation 
of these important signatures of QGP, there is a need for investigating 
novel signals exploring qualitatively non-trivial features of the QGP
phase and/or the quark-hadron phase transition.

   With this view we focus on the non-trivial vacuum structure of
the QGP phase which arises when one uses the expectation 
value of the Polyakov loop $l(x)$ as the order parameter for the
confinement-deconfinement phase transition \cite{lary}. This order 
parameter transforms non-trivially under the center $Z(3)$ of the color 
SU(3) group and is non-zero above the critical temperature $T_c$. This 
breaks the global $Z(3)$ symmetry spontaneously above $T_c$, while
the symmetry is restored below $T_c$ in the confining phase where
this order parameter vanishes. 
In the QGP phase, due to spontaneous breaking of the discrete $Z(3)$
symmetry, one gets domain walls (interfaces) which interpolate between
different $Z(3)$ vacua. The properties and physical consequences of 
these $Z(3)$ interfaces have been discussed in the literature\cite{zn}. 
It has been suggested  that these 
interfaces should not be taken as physical objects in the Minkowski 
space \cite{smlg}. Similarly, it has also been subject of discussion
whether it makes sense to talk about this $Z(3)$ symmetry in the presence
of quarks \cite{qurk1}. However, we will follow the approach where
the presence of quarks is interpreted as leading to explicit breaking 
of $Z(3)$ symmetry, lifting the degeneracy of different $Z(3)$ vacua 
\cite{qurk2,psrsk,psrsk2,z3lnr}. Thus, with quarks, even planar $Z(3)$ 
interfaces do not remain static and move away from the region with the 
unique true vacuum.  Our main discussion will be for the pure gauge 
theory which we discuss first. Later we will briefly comment on the 
situation with quarks. A detailed study with inclusion of quark effects 
is postponed for a future work.
 
  In earlier works some of us had shown that there are novel topological
string defects which form at the intersection of the three $Z(3)$
interfaces. These strings are embedded in the QGP phase and their
cores consist of the confining phase. Structure of these strings
and interfaces were discussed in these earlier works \cite{z3str}.
It was also shown that reflection of quarks from collapsing 
$Z(3)$ interfaces can lead to large scale baryon inhomogeneities in 
the early universe \cite{znb}. This effect is also utilized to
argue for $P_T$ enhancement of quarks of heavy flavor (consecutively
corresponding hadrons) in relativistic heavy-ion collisions
\cite{apm}.

  In this paper we carry out numerical simulation of formation of
these $Z(3)$ interfaces and associated strings at the initial 
confinement-deconfinement 
transition which is believed to occur during the pre-equilibrium
stage in relativistic heavy-ion collision experiments. For the
purpose of numerical simulation we will model this stage as
a quasi-equilibrium stage with an effective temperature which
first rises (with rapid particle production) to a maximum 
temperature $T_0 > T_c$, where $T_c$ is the critical temperature
for the confinement-deconfinement phase transition, and then
decreases due to continued plasma expansion. 

 We use the effective potential for the Polyakov loop expectation
value $l(x)$ as proposed by Pisarski \cite{psrsk,psrsk2} to 
study the confinement-deconfinement (C-D) phase transition. Within
this model, the C-D transition is weakly first order. Even though
Lattice results show that quark-hadron transition is most likely
a smooth cross-over at zero chemical potential, in the present work 
we  will use this first order transition model to discuss the dynamical
details of quark-hadron transition. One reason for this is that our 
study is in the context of relativistic heavy-ion  collision experiments 
(RHICE) where the baryon chemical potential is not zero. For not too 
small values of the chemical potential, the quark-hadron phase transition 
is expected to be of first order, so this may be the case relevant for
us any way (especially when collision energy is not too high). 
Further, our main interest is in determining the structure 
of the network of Z(3) domain walls and strings resulting during the 
phase transition. These objects will form irrespective
of the nature of the transition, resulting entirely from the finite
correlation lengths in a fast evolving system, as shown by 
Kibble \cite{kbl}. The Kibble mechanism was first proposed for the 
formation of topological defects in the context of the early universe
\cite{kbl}, but is now  utilized extensively for discussing topological 
defects production in a wide variety of systems from condensed matter 
physics to cosmology \cite{zrk}. Essential ingredient of the
Kibble mechanism is the existence of uncorrelated domains of the order
parameter which result after every phase transition occurring in finite
time due to finite correlation length. A first order transition allows
easy implementation of the resulting domain structure especially when
the transition proceeds via bubble nucleation. With this view, we use the
Polyakov loop model as in \cite{psrsk,psrsk2} to model the phase transition
and confine ourselves with temperature/time ranges so that the first 
order quark-hadron transition proceeds via bubble nucleation. 

 The $Z(3)$ wall network and associated strings (as mentioned above)
formed during this early confinement-deconfinement phase transition
evolve in an expanding plasma with decreasing temperature. Eventually
when the temperature drops below the deconfinement-confinement
phase transition temperature $T_c$, these $Z(3)$ walls and associated 
strings will melt away. However they may leave their signatures in the
form of extended regions of energy density fluctuations (as well
as $P_T$ enhancement of heavy-flavor hadrons \cite{apm}). We make
estimates of these energy density fluctuations which can be
compared with the experimental data. Especially interesting
will be to look for extended regions of large energy densities
in space-time reconstruction of hadron density (using hydrodynamic
models).  In our model, we expect energy 
density fluctuations  in event averages (representing high energy
density regions of domain walls/strings), as well as event-by-event
fluctuations as the number/geometry of domain walls/strings and even
the number of QGP bubbles, varies from one event to the other.
A detailed analysis of energy fluctuations, especially the
event-by-event fluctuations, is postponed for a future work.

 We also determine the distribution/shape of $Z(3)$ wall network and 
its evolution. In particular, our results provide an estimate of
domain wall velocities (for the situations studied) to range from 
0.5 to 0.8. These results provide  crucial ingredients for a detailed 
study of the effects of collapsing $Z(3)$ walls on the $P_T$ enhancement 
of heavy flavor hadrons \cite{znb,apm} in RHICE. We emphasize that the 
presence of $Z(3)$ walls and string may not only provide
a qualitatively new signature for the QGP phase in these experiments,
it may provide the first (and may be the only possible) laboratory
study of such topological objects in a relativistic quantum
field theory system.

 The paper is organized in the following manner. In section II, we 
discuss the Polyakov loop model of confinement-deconfinement phase 
transition. We describe the effective potential proposed by Pisarski 
\cite{psrsk} and discuss the structure of the $Z(3)$ walls and associated 
strings. In section III, we present the physical picture of the 
formation of these $Z(3)$ walls and associated strings in the 
confinement-deconfinement phase transition via the Kibble 
mechanism which provides a general framework for the production of 
topological defects in symmetry breaking phase transitions.
We confine ourselves with temperature/time
ranges so that the first order transition (in the Pisarski model)
proceeds via bubble nucleation. The other possibility of
spinodal decomposition is of completely different nature and we will 
present it in a future work. (We mention here that a simulation of 
spinodal decomposition in Polyakov loop model has been carried out in
ref. \cite{dumitru}, where fluctuations in the Polyakov loop
are investigated in detail. In comparison, the main focus of our 
work is on the topological objects, Z(3) walls, and strings, and 
energy density fluctuations resulting therefrom.)
In section IV, we discuss the calculation of
the profile of the critical bubble using bounce technique \cite{clmn}
and also the estimates of nucleation rates for bubbles for the 
temperature/time range relevant for RHICE. Section V presents the 
numerical technique of simulating the phase transition via random 
nucleation of bubbles. In section VI we discuss the issue of
the effects of quarks in our model. Section VII presents the results of 
the numerical simulations. Here we discuss distribution of $Z(3)$ 
wall/string network formed due to coalescence of QGP bubbles 
(in a confining background). We also calculate the energy density 
fluctuations associated with $Z(3)$ wall network and strings. We  discuss
evolution of these quantities with changing temperature via Bjorken's
hydrodynamical model. In section VIII we discuss possible experimental
signatures resulting from the presence of $Z(3)$ wall network and
associate strings. Section IX presents conclusions.

\section{THE POLYAKOV LOOP MODEL}

 We first discuss the case of  pure SU(N) gauge theory. We will later 
discuss briefly the case with quarks. The order parameter for 
the confinement-deconfinement (C-D) phase transition is the expectation
value of the Polyakov loop $\it l(x)$ which is defined as

\begin{equation}
\it l(x) =(1/N)tr(P exp(ig\int^\beta_0 A_0(x,\tau)d\tau))
\end{equation}

Here P denotes the path ordering, g is the gauge coupling,
and $\beta=1/T$, where T is the temperature. $A_0(x,\tau) \equiv
A_0^a(x,\tau)T_a$, $T_a$ being the generators of SU(N) in the fundamental
representation, is the time component of the vector potential at spatial 
position $\it {x}$ and Euclidean time $\tau$. Under a global Z(N) 
symmetry transformation, $\it l(x)$ transforms as

\begin{equation}
\it l(x) \rightarrow {exp(2{\pi}in/N)\it l(x)}, \: n = 0,1,...(N-1)
\end{equation}

 The expectation value of $l(x)$ is related to $e^{-\beta F}$ where
$F$ is the free energy of an infinitely heavy test quark. 
For temperatures below $T_c$, in the confined phase, the expectation value
of Polyakov loop  is zero corresponding to the infinite 
free energy of an isolated test quark. (Hereafter, we will use the
same notation $l(x)$ to denote the expectation value of the Polyakov
loop.)  Hence the Z(N) symmetry is restored below $T_c$. Z(N) symmetry 
is broken spontaneously above $T_c$ where $l(x)$ is non-zero  
corresponding to the finite free energy of the test quark. For QCD, 
$N = 3$, and we take the effective theory for the Polyakov loop as 
proposed by Pisarski \cite{psrsk,psrsk2}. The effective Lagrangian 
density is given by

\begin{equation}
L={N\over g^2} |{\partial_\mu\it l}|^2{T^2}- V(\it l)
\end{equation}

$V(\it l)$ is the effective potential for the Polyakov loop

\begin{equation}
 V(\it l)=({-b_2\over2} |\it l|^2- {b_3\over 6}(\it l^3+(\it l^ \ast)^
 3)+\frac{1}{4}(|\it l|^2)^2){b_4{T^4}}
\end{equation}

The values of various parameters are fixed to reproduce the lattice 
results \cite{scav,latt} for pressure and energy density of pure SU(3) 
gauge theory. We make the same choice and give those values below. 
The coefficients $b_3$ and $b_4$ have been taken as, $b_3=2.0$ and 
$b_4=0.6016$. We will take 
the same value of $b_2$ for real QCD (with three massless quark flavors), 
while the value of $b_4$ will be rescaled by a factor of 47.5/16 to account 
for the extra degrees of freedom relative to the degrees of freedom of pure 
SU(3) gauge theory \cite{scav}. The coefficient $b_2$ is temperature dependent 
\cite{scav}. $b_2$ is taken as,
$b_2(r) = (1-1.11/r)(1+0.265/r)^2(1+0.3/r)^3-0.487$, where r is 
taken as T/$T_c$. With the coefficients chosen as above, the expectation 
value of order parameter approaches to 
$\it x =b_3/2+\frac{1}{2} \sqrt{b_3^2+4b_2(T=\infty)}$ for temperature 
$T \rightarrow \infty$. As in \cite{psrsk2}, we use 
the normalization such that the expectation value of order parameter $\it l_0$  
goes to unity for temperature $T \rightarrow \infty$. Hence the fields and 
the coefficients in $V(l)$ are rescaled as $ \it l \rightarrow l/x$, 
$b_2(T)\rightarrow b_2(T)/x^2$, $b_3\rightarrow b_3/x$ and 
$b_4\rightarrow b_4x^4$ to get proper normalization of $\it l_0$.

For the parameters chosen as above, the value of $T_c$ is taken to be
182 MeV. We see that the $b_3$ term in Eq.(4) gives a cos(3$\theta$) term, 
leading to Z(3) degenerate vacua structure. Here, the shape of the 
potential is such that there exists a metastable vacuum upto a temperature 
$\sim$ 250 MeV. Hence first order transition via bubble nucleation is 
possible only upto $T = 250$ MeV. We show 
the plot of V$(\it l)$ in $\theta=0$ direction in Fig.(1a) for a 
value of temperature T = 185 MeV. This shows the metastable vacuum at 
$\it l = 0$. Fig.(1b) shows the structure of vacuum by plotting V($\it l$) 
as a function of $\theta$ for fixed $\it l_0$, where $l_0$ is the vacuum 
expectation value of $V(\it l)$ at $T = 185$ MeV.

\begin{figure*}[!hpt]
\begin{center}
\leavevmode
\includegraphics[width=0.3\textwidth]{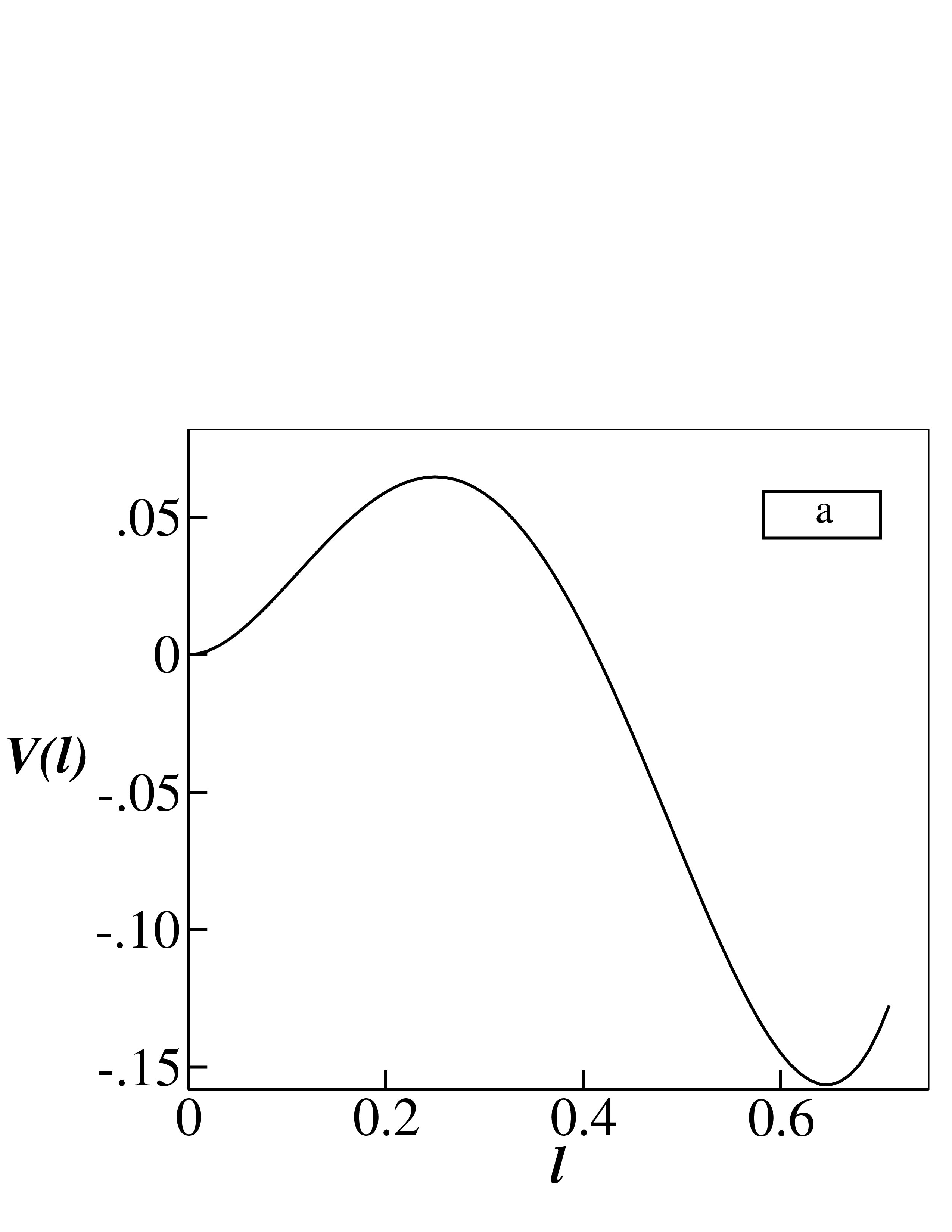} \\
\includegraphics[width=0.3\textwidth]{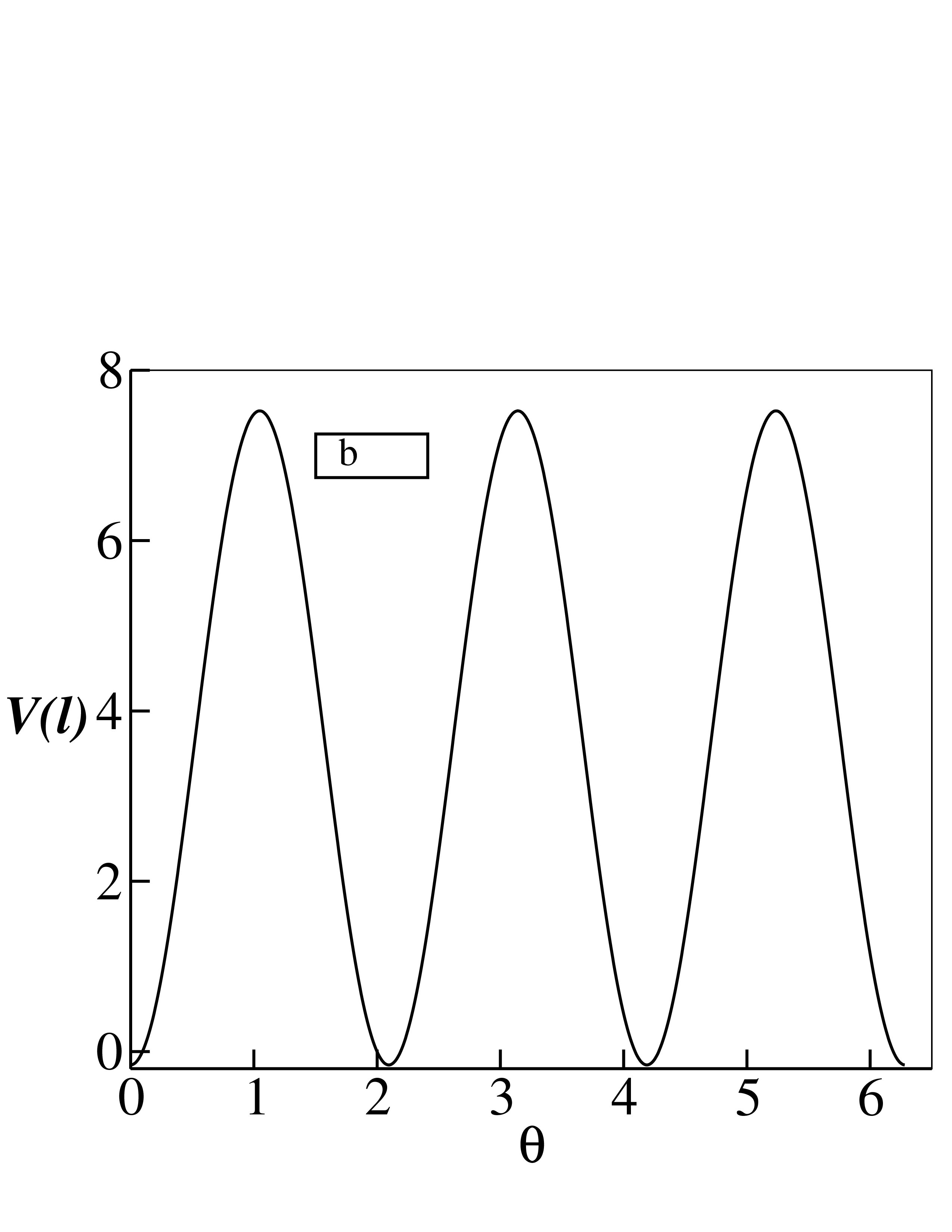}
\end{center}
\caption{}{(a) Plot of  $V(l)$ in $\theta = 0$ direction for $T = 185 MeV$
showing the metastable vacuum at $l = 0$.  In (a) and (b), plots of $V$ 
are given in units of $T_c^4$. The value of critical temperature is taken 
to be $T_c \simeq 182 MeV$. The $Z(3)$ structure of the vacuum  can be 
seen in (b) in the plot of the potential $V(l)$ as a function of $\theta$ 
for fixed $|l| = l_0$. Here, $l_0$ corresponds to the absolute minimum of 
$V(l)$.}
\label{Fig.1}
\end{figure*}

In Fig.(1b), the three degenerate vacua are separated by large barrier 
in between them. While going from one vacuum to another vacuum, the field 
configuration is determined from the field equations. For the
case of degenerate vacua, there are time independent solutions
which have planar symmetry. These solutions are called domain wall.
For the non-degenerate case, as will be appropriate for the case
when quarks are included as dynamical degrees of freedom in discussing
the quark hadron transition, the solutions of the interfaces separating
these vacua will be similar to the bounce solutions \cite{clmn}, though 
the standard bounce techniques need to be extended for the case of complex 
scalar field. The resulting planar domain wall solutions will not be static.
As mentioned above, we will be neglecting such effects of quarks, and 
hence will discuss the case of degenerate vacua only. Later, in 
section VI, we will examine the justification of using the 
approximation of neglecting quark effects.
 
In physical space, after the phase transition, regions with different 
Z(3) vacua  are separated by domain walls. Inside a domain wall, $|l|$ 
becomes very small as $T \rightarrow T_c$ 
\cite{z3str}. In ref.\cite{z3str} the intersection of the three Z(3) 
domain walls was considered and, using topological arguments, it was
then shown that at the line-like intersection of these
interfaces, the order parameter $l(x)$ should vanish. This leads to 
a topological string configuration with core of the string being in 
the confining phase. Properties of these new string configurations
were determined in \cite{z3str} using the model of Eq.(3). This
string configuration is interesting, especially when we note
that such a string has exactly reverse physical behavior compared to
the standard QCD string. The QCD string exists in the confining phase,
connecting quarks and antiquarks, or forming baryons, glueballs etc.
Inside the QCD string,  the core region is expected to behave 
as  a deconfined region. In contrast the string discussed in
\cite{z3str}, arising at the intersection of Z(3) walls, exists in 
the high temperature deconfined phase. Its core is characterized by 
restored $Z(3)$ symmetry, implying that it is in the confined phase. 
To differentiate it with the standard QCD string, this new string
structure was called as {\it the QGP string} in ref.\cite{z3str}. 
It is also important to note that although the standard
QCD string breaks by creating quark-antiquark pairs, the QGP string
cannot break as it originates from topological arguments. This QGP
string, thus should either form closed loops, or it should end at the
boundary separating the deconfined phase from the confined phase.
Its structure is very similar to certain axionic strings discussed
in the context of early universe \cite{axion}. Note that as these 
strings contain the confining phase (with $l = 0$) in the core, 
while they are embedded in the QGP phase, a transition
from deconfining phase to the confining phase, in the presence of
such strings may begin from regions near the strings. Similarly, the
presence of domain walls may lead to heterogeneous bubble nucleation
in a first order quark-hadron transition.
We will be studying the formation of Z(3) domain walls and these
QGP strings in the initial confinement-deconfinement phase transition,
and their subsequent evolution.

\section{DOMAIN WALL  AND STRING FORMATION VIA KIBBLE MECHANISM}

 We now briefly describe the physical picture of the 
formation of topological defects via the Kibble mechanism.
Kibble first gave a detailed theory of the formation of topological 
defects in symmetry breaking phase transitions in the 
context of the early universe \cite{kbl}. Subsequently it was realized that
the basic physics of Kibble mechanism is applicable to every
symmetry breaking transition, from low energy physics of 
condensed matter systems to high energy physics relevant for
the early universe \cite{zrk}. 

 Basic physics of the Kibble mechanism can be described as follows.
After a spontaneous symmetry breaking phase transition, the physical 
space consists of regions, called domains. In each domain the 
configuration of the order parameter field can be taken as nearly 
uniform while it varies randomly from one domain to another. In 
a numerical simulation where the phase transition is modeled to
implement the Kibble mechanism, typically the physical region is
divided in terms of elementary domains of definite geometrical shape.
The order parameter is taken to be uniform within the domain and
random variations of order parameter field within the vacuum manifold
are allowed from one domain to the other. The order parameter field 
configuration in between domains is assumed to be such that the 
variation of the order parameter field is minimum on the vacuum 
manifold (the so-called geodesic rule). With this simple construction,
topological defects arise at the junctions of several domains if
the variation of the order parameter in those domains traces a
topologically non-trivial configuration in the vacuum manifold.
See, ref. \cite{vlnkn} for a detailed discussion of this approach. 

We, however, will follow a more detailed simulation as in 
ref.\cite{ajit} where the Kibble mechanism was implemented in the
context of a first order transition. Bubbles of true
vacuum (determined from the bounce solution) were randomly nucleated
in the background of false vacuum. Each bubble was taken to have the
uniform orientation of the order parameter in the vacuum manifold,
while the order parameter orientation varied randomly from one
bubble to another. This provided the initial seed domains, as needed for
the Kibble mechanism. Evolution of this initial configuration via
the field equation led to expansion of bubbles which eventually
coalesce and lead to the formation of topological defects at
the junctions of bubbles when the order parameter develops appropriate
variation (winding) in that region. Important thing is that in this
case one does not need to assume anything like the geodesic rule.
As different bubbles come into contact during their expansion, 
the value of the order parameter in the intermediate region is
automatically determined by the field equations.  

An important aspect of the Kibble mechanism is that it does not crucially 
depend on the dynamical details of the phase transition. Although the domain 
size depends on the dynamics of phase transitions, the defect number density
(per domain) and type of 
topological defects produced via the Kibble mechanism depends only on the 
topology of order parameter space and spatial dimensions. If the vacuum
manifold M has disconnected components, then domain walls form. If it is 
multiply connected (i.e., if M contains unshrinkable loops), strings 
will form. When M contains closed two surfaces which cannot be shrunk 
to a point, then monopoles will form in three dimensional physical space.  
In our case  domain walls and string network will be produced in the QGP 
phase. As discussed above, domain walls arise due to interpolation of field 
between different Z(3) vacua. At the intersection of these interfaces, 
string is produced. 

\section{CRITICAL BUBBLE PROFILE AND NUCLEATION PROBABILITY}

 As we explained in the introduction, we will study the Z(3) domain
wall and string formation with the first order transition model
given by Eq.(3), such that the transition occurs via bubble nucleation.
The semiclassical theory of decay of false vacuum at zero temperature
has been given in ref. \cite{clmn}, and the finite temperature extension
of this theory was given in ref. \cite{linde}. The process of barrier
tunneling leads to the appearance of bubbles of the new phase.
Resulting bubble profile is determined using the bounce solution for
the false vacuum decay which we will discuss below.

 First we note general features of the dynamics of a standard first order 
phase transition at finite temperature via bubble nucleation. A region 
of true vacuum, in the form of a spherical bubble, appears in the 
background of false vacuum. The creation of bubble leads to the change 
in the free energy of the system as,

\begin{equation}
{\it F}(R) = {\it F}_s + {\it F}_v = 
4\pi R^2 \sigma- {4\pi \over 3} R^3 \eta
\end{equation}

where R is the radius of bubble, ${\it F_s}$  is the surface
energy contribution and ${\it F_v}$ is the volume energy contribution.
$\eta$ is the difference of the free energy between the false vacuum and 
the true vacuum and $\sigma$ is the surface tension which can be determined 
from the bounce solution. 

A bubble of size $R$ will expand or shrink depending on which
process leads to  lowering of the free energy given above.
The bubbles of very small sizes will shrink to nothing since 
surface energy dominates. If the radius of bubble exceeds the critical
size  $R_c = {2 \sigma \over \eta}$, it will expand and lead to the 
transformation of the metastable phase into the stable phase. 

 Eq.(5) is useful for the so-called {\it thin wall} bubbles where there
is a clear distinction between the surface contribution to the free energy
and the volume contribution. For the temperature/time relevant for our case
in relativistic heavy-ion collisions this will not be the case. Instead
we will be dealing with the {\it thick wall} bubbles where surface and
volume contributions do not have clear separations. We will determine
the profiles of these thick wall bubbles numerically following the
bounce technique \cite{clmn}.  

 First we note that for the effective potential in Eq.(4), the barrier 
between true vacuum and false vacuum vanishes at temperatures above
about  250 MeV. So first order transition via bubble nucleation is 
possible only within the temperature range of $T_c \simeq 182$ MeV - 250 MeV. 
Above $T \simeq$ 250 MeV, spinodal decomposition will take place due to 
the roll down of field. Implementation of the dynamics of phase transition
via spinodal decomposition is of completely different nature, and we hope
to discuss this in a future work.

 As we are discussing the initial confinement-deconfinement transition 
in the context of RHICE, clearly
the discussion has to be within the context of longitudinal
expansion only, with negligible effects of the transverse expansion.
However, Bjorken's longitudinal scaling model \cite{bjorken} cannot
be applied during this pre-equilibrium phase, even with the assumption 
of quasi-equilibrium (as discussed above), unless one includes a heat
source which could account for the increase of effective temperature
during this phase to the maximum equilibrium temperature $T_0$. As
indicated above, this heat source can be thought of as representing
the rapid particle production (with subsequent thermalization) during
this early phase. We will not attempt to model such a source here.
Instead, we will simply use the field equations resulting from
Bjorken's longitudinal scaling model 
for the evolution of the field configuration for the entire simulation,
including the initial pre-equilibrium phase from $\tau = 0$ to $\tau = 
\tau_0$. The heating of the system  until $\tau = \tau_0$ will be 
represented by the increase of the temperature upto $T = T_0$.
Thus, during this period, the energy density and temperature evolution
will not obey the Bjorken scaling equations \cite{bjorken}. After
$\tau_0$, with complete equilibrium of the system, the temperature 
will decrease according to the equations in the Bjorken's
longitudinal scaling model. 

   We will take the longitudinal expansion to only represent the 
fact that whatever bubbles will be nucleated, they get stretched into
ellipsoidal, and eventually cylindrical, shapes during the longitudinal 
expansion (ignoring the boundary effects in the longitudinal direction). 
The transverse expansion of the bubble should then proceed according to
relative pressure difference between the false vacuum and the true
vacuum as in the usual theory of first order phase transition. We will 
neglect transverse expansion for the system, and focus on the mid 
rapidity region. With this picture in mind, we will work with effective 
2+1 dimensional evolution of the field configuration, (neglecting the 
transverse expansion of QGP). However, for determining the
bubble profile and the nucleation probability of bubbles, one must 
consider full 3+1 dimensional case  as bubbles are nucleated with 
full 3-dimensional profiles in the physical space. It will turn
out that the bubbles will have sizes of about 1 - 1.5 fm radius.
Taking the initial collision region during the pre-equilibrium
phase also to be of the order of 1-2 fm in the longitudinal direction,
it looks plausible that the nucleation of 3-dimensional bubble 
profile as discussed above may provide a good approximation. 
Of course the correct thing will be to consider the bounce solutions for
rapidly longitudinally expanding plasma, and we hope to return
to this issue in some future work.  

  We neglect transverse expansion in the present work, which is
a good approximation for the early stages when wall/string network
forms. However, this will not be a valid approximation for later
stages, especially when temperature drops below $T_c$ and wall/string
network melts. The way to account for the transverse expansion in
the context of our simulation will be to take a lattice with much larger
physical size than the initial QGP system size, and allow free boundary 
conditions for the field evolution at the QGP system boundary (which will 
still be deep inside the whole lattice). This will allow the freedom for 
the system to expand in the transverse direction automatically. With
a suitable prescription of determining temperature from local energy 
density (with appropriate account of field contributions and expected
contribution from a plasma of quarks and gluons) in a self consistent 
manner, the transverse expansion can be accounted for in this
simulation. We hope to come back to this in a future work. 

 Let us consider the effective potential in Eq.(4), at a temperature
such that there is a barrier between the true vacuum and the three Z(3) 
vacua. An example of this situation is shown in Fig.1 for the case 
with $T = $ 185 MeV. The initial system
(of nucleons) was at zero temperature with the order parameter l(x) = 0, 
and will be superheated as the temperature rises above the critical 
temperature. It can then tunnel through the barrier to the true vacuum,
representing the deconfined QGP phase. At zero temperature, the
tunneling probability can be calculated by finding the bounce solution
which is a solution of the 4-dimensional Euclidean equations of motion 
However, at finite temperature, this 4-dimensional theory will
reduce to an effectively 3 Euclidean dimensional theory if the temperature 
is sufficiently high, which we will take to be the case.

For this finite temperature case, the tunneling probability per unit 
volume per unit time in the high temperature approximation is 
given by \cite{linde} (in natural units)

\begin{equation}
 \Gamma = A~e^{-S_3(l)/T} 
\end{equation}

where $S_3(l)$ is the 3-dimensional Euclidean action for the  Polyakov
loop field configuration that satisfies the classical Euclidean equations 
of motion. The condition for the high temperature approximation to be valid 
is that $T >> r_0^{-1}$, where $r_0$ is the radius of the critical 
bubble in 3 dimensional Euclidean space. The values of temperature 
for our case (relevant for bubble nucleation) will be  above $T = T_c
= 182$ MeV. As we will see, the bubble radius will be larger than 1.5
fm ( $\sim$ (130 MeV)$^{-1}$) which justifies our use of high temperature 
approximation to some extent. The determination of the pre-exponential 
factor is a non-trivial issue and we will discuss it below. The dominant 
contribution to the exponential term in $\Gamma$ comes from the 
least action $O(3)$ symmetric configuration which is a solution 
of the following equation (for the Lagrangian in Eq.(3)).

\begin{equation}
{d^2 l\over dr^2} +{2\over r}{d l\over dr} 
={g^2\over {2NT^2}}V^\prime(\it l)
\end{equation} 

\noindent where $r \equiv r_E = \sqrt{{\vec x}^2 + t_E^2}$, subscript E
denoting the coordinates in the Euclidean space.

The boundary conditions imposed on $l$  are

\begin{equation}
l =0 ~~{\rm as}~~ r\rightarrow\infty \qquad {\rm and} \qquad
{d l \over dr}=0 ~~{\rm at}~~ r=0
\end{equation}

 Bounce solution of Eq.(7) can be analytically obtained in the {\it thin wall}
limit where the difference in the false vacuum and the true vacuum energy
is much smaller than the barrier height. This situation will occur for
very short time duration near $T = T_c$ for the effective potential
in Eq.(4). However, as the temperature is
rapidly evolving in the case of RHICE, there will not be enough time
for nucleating such large bubbles (which also have very low nucleation
rates due to having large action). Thus the case relevant for us is
that of thick wall bubbles whose profile has to be obtained by 
numerically solving Eq.(7).

 As we have mentioned earlier, in the high temperature approximation 
the theory effectively  becomes 3 (Euclidean)  dimensional. For a theory 
with one real scalar field in three Euclidean dimensions the 
pre-exponential factor arising in the nucleation rate of critical 
bubbles has been estimated, see ref. \cite {linde}. The pre-exponential
factor obtained from \cite {linde} for our case becomes

\begin{equation}
 A =  T^4 \left({S_3(l) \over 2\pi T}\right)^{3/2}
\end{equation}

 It is important to note here that the results of \cite {linde} were 
for a single real scalar field and one of the crucial ingredients 
used in \cite {linde} for calculating the pre-exponential factor was
the fact that for a bounce solution the only light modes contributing to
the determinant of fluctuations were the deformations
of the bubble perimeter. Even though we are discussing the case of 
a complex scalar field $l(x)$, this assumption may still hold as we
are calculating the tunneling from the false vacuum to one of the
Z(3) vacua (which are taken to be degenerate here as we discussed
above). This assumption may need to be revised when 
light modes e.g. Goldstone bosons are present which then also have 
to be accounted for in the calculation of the determinant. 

 A somewhat different approach for the pre-exponential factor in Eq.(6)
is obtained from the nucleation rate of  bubbles per unit volume 
for a liquid-gas phase transition as given in ref. \cite{langer,csern}, 

\begin{equation}
\Gamma = { \kappa \over {2\pi}}{\Omega_0} e^{-\Delta F/T}
\end{equation}

Here $\kappa$ is the dynamical prefactor which determines the exponential 
growth rate of critical droplets. $\Omega_0$ is a statistical prefactor which 
measures the available phase space volume. The exponential term is the
same as in Eq.(6) with $\Delta F$ being the change in the free energy of 
the system due to the formation of critical droplet. This is the same as
$S_3$ in Eq.(6). The bubble grows beyond the critical size when the latent 
heat is conducted away from the surface into the surrounding medium which 
is governed by thermal dissipation and viscous damping. For our case, in 
the general framework of transition from a hadronic system to the QGP phase, 
we will use the expression for the dynamical prefactor from ref.\cite{kpst}

\begin{equation}
\kappa  = {2\sigma \over (\Delta \omega)^2 R_c^3} 
(\lambda T + 2({4 \over 3} \eta + \zeta))
\end{equation}

 Here $\sigma$ is the surface tension of the bubble wall, 
$\Delta \omega$ is the difference in the enthalpy densities of the
QGP and the hadronic phases, $\lambda$ is thermal conductivity, $R_c$ is
the critical bubble radius, and $\eta$ and $\zeta$ are shear and bulk 
viscosities. $\zeta$ will be neglected as it is much smaller than $\eta$.
For $\lambda$ and $\eta$, the following parametrizations are used
\cite{kpst,daniel}.

\begin{equation}
\eta = \left({1700 \over T^2}\right) \left({n \over n_0}\right)^2 + 
\left({22 \over 1 + T^2/1000}\right)
\left({n \over n_0}\right)^{0.7} + {5.8 T^{1/2} \over 1 + 160/T^2} 
\end{equation}

\begin{equation}
\lambda = \left({0.15 \over T}\right) \left({n \over n_0}\right)^{1.4} 
+ \left({0.02 \over 1 + T^4/ (7 \times 10^6)}\right) 
\left({n \over n_0}\right)^{0.4} + {0.0225 T^{1/2} \over 1 + 160/T^2}
\end{equation}

 Here $n/n_0$ is the ratio of the baryon density of the system to
the normal nuclear baryon density, $T$ is in MeV, $\eta$ is in
MeV/fm$^2$c, and $\lambda$ is in c/fm$^2$. With this, the rate in 
Eq.(10) is in fm$^{-4}$. For the range of temperatures 
of our interest ($T > 160 MeV$), and for the low baryon density central
rapidity region  under consideration, it is the last $n$ independent 
term for both $\eta$ and $\kappa$ which dominates, and we will use 
these terms only for calculating $\eta$ and $\kappa$ for our case.   

For the statistical prefactor, we use the following expression \cite{kpst}

\begin{equation}
\Omega_0={{2\over{3\sqrt3}} \left({\sigma \over T}\right)^{3/2}
\left({R_c \over \xi_{had}}\right)^4}
\end{equation}

The correlation length in the hadronic phase, $\xi_{had}$, is expected
to be of order of 1 fm and we will take it to be 0.7 fm \cite{kpst}). 

 We will present estimates of the nucleation rates from Eq.(6) as well
as Eq.(10). One needs to determine the critical bubble profile and
its 3-dimensional Euclidean action $S_3$ (equivalently, $\Delta F$ in
Eq.(10)). We solve Eq.(7) using the fourth order Runge-Kutta method with 
appropriate boundary conditions (Eq.(8)), to get the profile of critical 
bubble \cite{ajit}. The critical bubble profiles (for the 3+1 dimensional
case) are shown in Fig.(2a) for 
different temperatures. The bubble size decreases as temperature 
increases, since the energy difference between true vacuum and false 
vacuum increases (relative to the barrier height) as temperature 
increases. We choose a definite 
temperature $T = 200$ MeV for the nucleation of bubbles, which is 
suitably away from $T_c$ to give acceptable bubble size and nucleation 
probabilities for the relevant time scale. Making $T$ larger (up to $T 
= 250$ MeV when the barrier disappears) leads to similar  bubble profile, 
and the nucleation probability of same order. 

\begin{figure*}[!hpt]
\begin{center}
\leavevmode
\includegraphics[width=0.4\textwidth]{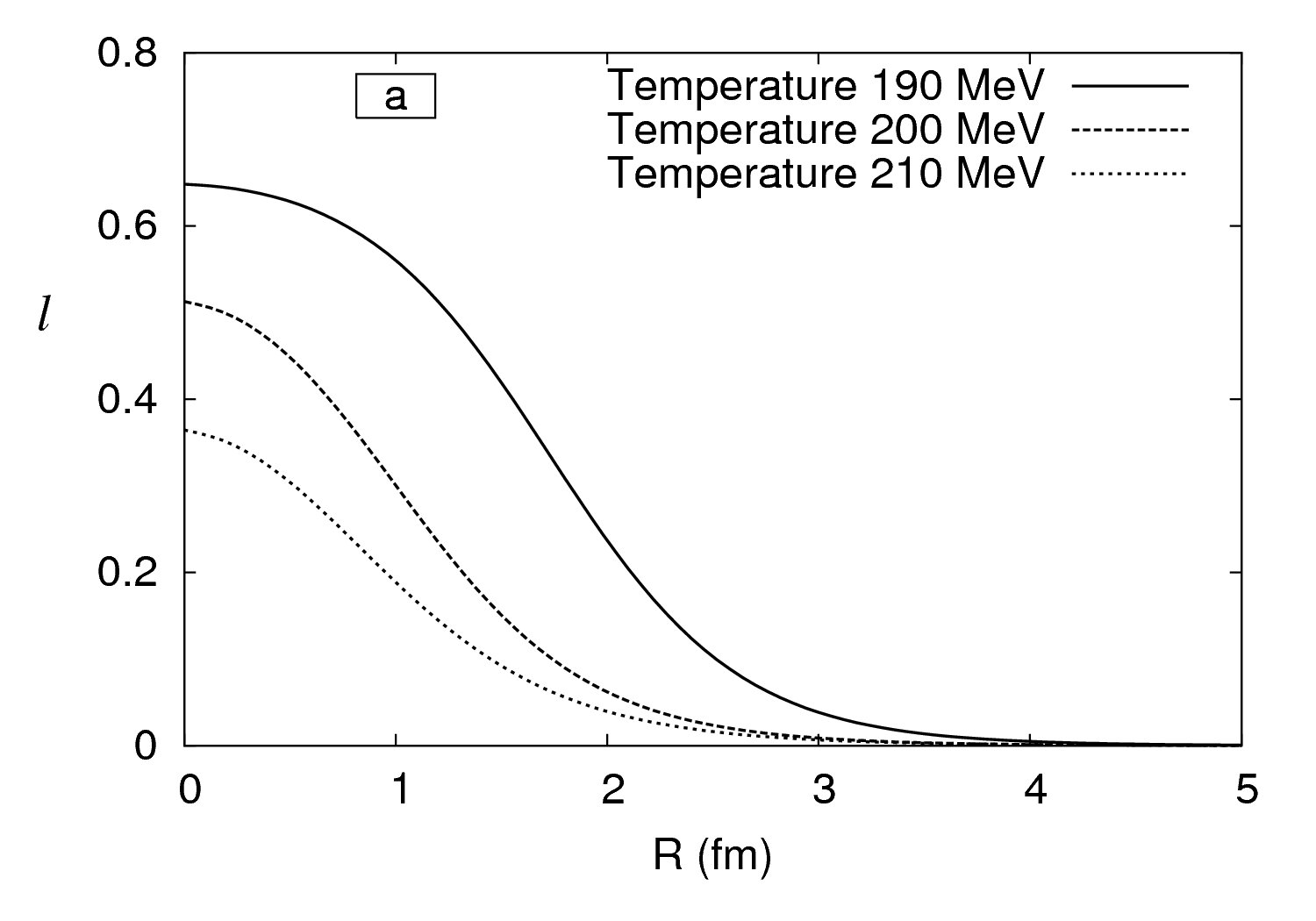} \\
\includegraphics[width=0.4\textwidth]{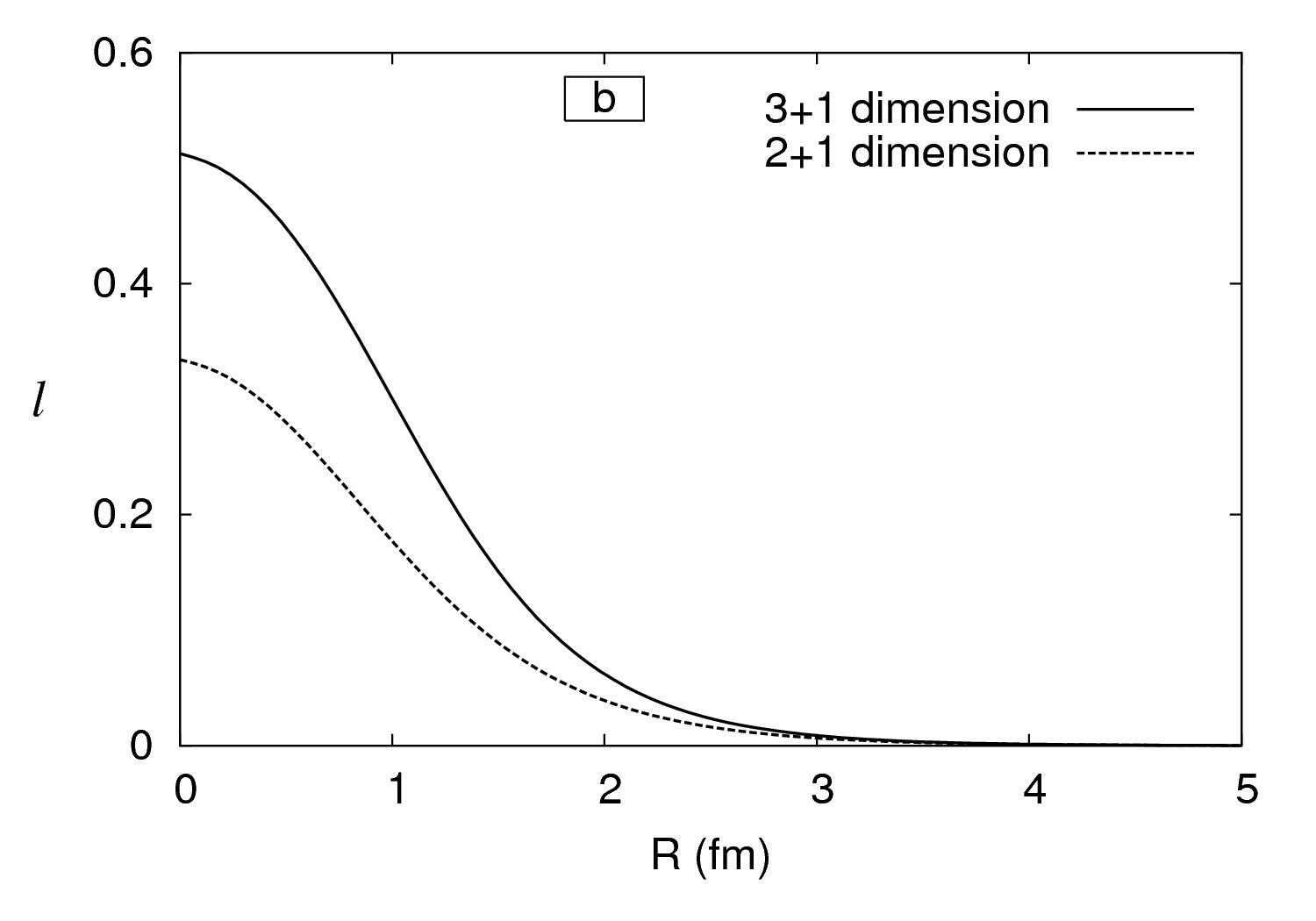} 
\end{center}
\caption{}{(a) Critical bubble profiles for different values
of the temperature. (b) Solid curve shows the critical bubble for 
the 3+1 dimensional case (which for finite temperature case becomes
3 Euclidean dimensional) for $T$ = 200 MeV  and the dotted curve
shows the same for 2+1 dimensional case (i.e. 2 Euclidean dimensions
for finite temperature case).}
\label{Fig.2}
\end{figure*}

  Recall that we are calculating bubble profiles using Eq.(7)
relevant for the 3+1 dimensional case, however, the field evolution
is done using 2+1 dimensional equations as appropriate for the
mid rapidity region of rapidly  longitudinally expanding plasma.
In Fig.(2b) solid curve shows the critical bubble for the 3+1 
dimensional case (which for finite temperature case becomes
3 Euclidean dimensional) for $T$ = 200 MeV  and the dashed curve
shows the same for 2+1 dimensional case (i.e. 2 Euclidean dimensions
for finite temperature case). It is clear that the
3 dimensional bubble is of supercritical size and will
expand when evolved with 2+1 dimensional equations. This avoids the 
artificial construction of suitable supercritical bubbles which 
can expand and coalesce as was done in ref.\cite{ajit}. (Recall
that for the finite temperature case, a bubble of exact critical size 
will remain static when evolved by the field equations. In a
phase transition, bubbles with somewhat larger size than the critical 
size expand, while those with smaller size contract.)

  For the bubble profile given by the solid curve in Fig.(2b), the value 
of the action $S_3(l)$ is about 240 MeV. Using Eq.(6) for
the nucleation rate, we find that
the nucleation rate of QGP bubbles per unit time per unit volume is 
of the order of $0.025 fm^{-4}$. The thermalization time for QGP phase
is of the order of 1 fm at RHIC (say, for Au-Au collision at 200 GeV 
energy). Hence the time available for the nucleation of QGP bubbles 
is at most about 1 fm. We take the region of bubble nucleation to
be of thick disk shape with the radius of the disk (in the transverse
direction) of about 8 fm and the thickness of the disk (in the
longitudinal direction) of about 1 fm. Total space-time volume 
available for bubble nucleation is then about 200 fm$^4$ (in practice,
less than this). For the case of Eq.(6), net number of bubbles is then 
equal to 5.  

For the case of the nucleation rate given by Eq.(10), one needs an estimate 
of the critical bubble size $R_c$ as well as bubble surface energy
$\sigma$, (along with other quantities like $\lambda$ etc. as given
by Eqs.(11)-(14)). Determination of $R_c$ is somewhat ambiguous here
as the relevant bubbles are thick wall bubbles as seen in Fig.(2).
Here there is no clear demarcation between the core region and the
surface region which could give an estimate of $R_c$.  Essentially,
there is no core at all and the whole bubble is characterized by
the overlap of bubble wall region. We can take, as an estimate for 
the bubble radius $R_c$ any value from 1 fm - 1.5 fm. It is important
to note here that this estimate of $R_c$ is only for the calculation
of nucleation rate $\Gamma$, and not for using the bubble profile
for actual simulation. When bubbles are nucleated in the background
of false vacuum with $l=0$, a reasonably larger size of the bubble
is used so that cutting off the profile at that radius does not
lead to computational errors and field evolution remains smooth.

  Once we have an estimate of $R_c$, we can then estimate the
surface tension $\sigma$ (which also is not unambiguous here) as
follows. With the realization that essentially there is no core
region for the bubbles in Fig(2), we say that the entire energy
of the bubble (i.e. the value of $S_3$) comes from the surface
energy. Then we write  

\begin{equation}
4 \pi R_c^2 \sigma = S_3
\end{equation}

 For $S_3$ = 240 MeV, we get $\sigma = 8 $  MeV/fm$^2$ if we take
$R_c$ = 1.5 fm. With $R_c$ = 1.0 fm, we get $\sigma$ = 20 MeV/fm$^2$.

 Number of bubbles expected can now be calculated for the case when 
nucleation rate is given by Eq.(10). We find number of bubbles to be 
about 10$^{-4}$ with $R_c$ taken as 1.5 fm. This is in accordance with 
the results discussed in ref. \cite{kpst}. The bubble number increases 
by about a factor 5 if $R_c$ is taken to be about 1 fm. Thus, with the 
estimates based on Eq.(10), bubble nucleation is a rare event for the time 
available for RHICE.  

  As we have mentioned above, for us the bubble nucleation on one
hand represents the possibility of actual dynamics of a first order 
transition, while on the other hand, it represents the generic properties 
of the domain structure arising from a C-D transition, which may
very well be a cross-over, occurring in a finite time.
With this view, and with various uncertainties in the determination 
of pre-exponential factors in the nucleation rate, we will consider 
a larger number of bubbles also and study domain wall and 
string production. First we will consider nucleation of 5 bubbles 
and then we will consider nucleation  of 9 bubbles to get a better 
network of domain walls and strings.

\section{NUMERICAL TECHNIQUES}

In our simulation critical bubbles are nucleated at a time when the
temperature $T$ crosses the value $T = 200$ MeV during the
initial stage between $\tau = 0$ to $\tau = \tau_0 = 1$ fm (during which
we have modeled the system temperature to increase linearly from 0 to
$T_0$). We take $T_0 = 400$ MeV so bubble nucleation stage is taken
to be at $\tau = 0.5 $ fm when $T$ reaches the value 200 MeV. Again, this is
an approximation since in realistic case bubbles will nucleate over
a span of time given by the (time dependent) nucleation rate, which could 
lead to a spectrum of sizes of expanding bubbles at a given time. However,
due to very short time available to complete the nucleation of QGP bubbles
in the background of confined phase, bubbles will have very
little time to expand  during the nucleation period of all bubbles 
(especially as initial bubble expansion velocity is zero). Thus it
is reasonable to assume that all the bubbles nucleate at the same time. 

After nucleation, bubbles are evolved by the time dependent equations of 
motion in the Minkowski space \cite{rndrp} as appropriate for Bjorken's
longitudinal scaling model.

\begin{equation}
{\ddot l_i} + {{\dot l_i} \over \tau} - {\partial^2 l_i \over 
\partial x^2} - {\partial^2 l_i \over \partial y^2} =
{-\partial V(l) \over {\partial l_i}}, \qquad   i=1,2
\end{equation}      

with ${\dot l} = 0$ at $\tau$ = 0. Here $l = l_1+i l_2$, and dot 
indicates derivative with respect to the proper time $\tau$.

The bubble evolution was numerically implemented by a stabilized leapfrog
algorithm of second order accuracy both in space and in time with the
second order derivatives of $l_i$ approximated by a diamond shaped grid.
Here we follow the approach described in \cite{ajit}
to simulate the first order transition. We need to nucleate several
bubbles randomly choosing the corresponding Z(3) vacua for each bubble. 
This is done by randomly choosing the location of the center of each 
bubble with some specified probability per unit time per unit volume. 
Before nucleating a bubble, it is checked if the relevant region is 
in the false vacuum (i.e. it does not overlap with some other 
bubble already nucleated). In case there is an overlap, the nucleation of the 
new bubble is skipped. The orientation of $l$ inside each bubble is 
taken to randomly vary between the three Z(3) vacua.

For representing the situation of relativistic heavy ion collision 
experiments, the simulation of the phase transition
is carried out by nucleating bubbles on a square lattice with physical 
size of 16 fm within a circular boundary (roughly the Gold nucleus size). 
We use fixed boundary condition, free boundary condition, as well as 
periodic boundary condition for the square lattice. To minimize the 
effects of boundary (reflections for fixed boundary, mirror reflections 
for periodic boundary conditions), we present results for free boundary
conditions (for other cases, the qualitative aspects of our results
remain unchanged). Even for free boundary conditions, spurious partial 
reflections occur, and to minimize these effects we use a thin
strip (of 10 lattice points) near each boundary where extra dissipation
is introduced.

We use 2000 $\times$ 2000 lattice. For the physical size of 16 fm, we
have $\Delta x$ = 0.008 fm. To satisfy the Courant stability criteria, 
we use $\Delta t = \Delta x/\sqrt 2$, as well as $\Delta t = 0.9 
\Delta x/\sqrt 2$, (which we use for the results presented in the 
paper). For Au-Au collision at 200 GeV, the thermalization is expected 
to happen within 1 fm time. As mentioned above, in this 
pre-equilibrium stage, we model the system as being in a quasi-equilibrium
stage with a temperature which increases linearly with time (for simplicity). 
The temperature of the system is taken to reach upto 400 MeV in 1 fm time,
starting from $T$ = 0. After $\tau = \tau_0 = 1$ fm, the temperature 
decreases due to continued longitudinal expansion, i.e.

\begin{equation}
T(\tau) = T(\tau_0) \left({\tau_0 \over \tau}\right)^{1/3}
\end{equation}

 The stability of the simulation
is checked by checking the variation of total energy of the system during
the evolution. The energy fluctuation remains within few percent, with no
net increase or decrease in the energy (for fixed and periodic boundary
conditions, and without the dissipative $\dot l$ term in Eq.(16)) showing 
the stability of the simulation.

The bubbles grow and eventually start coalescing, leading to a domain 
like structure. Domain walls are formed between regions 
corresponding to different Z(3) vacua, and strings
form at junctions of Z(3) domain walls. Recall that the domain wall network 
is formed here in the transverse plane, appearing as curves. These are
the cross-sections of the walls which are formed by elongation 
(stretching) of these curves in the longitudinal direction into sheets. 
At the intersection of these walls, strings form. In the transverse 
plane, these strings looks like vortices, which  will be  elongated 
into strings in the  longitudinal direction.

\section{Effects of quarks}

 We now discuss the effects of quarks. As we mentioned above, we will
follow the approach where the presence of quarks is interpreted as 
leading to explicit breaking of the Z(3) symmetry, lifting the degeneracy 
of different Z(3) vacua \cite{qurk2,psrsk,psrsk2}. This has important
effects in the context of our model. First of all, different vacua
having different energies implies different nucleation rates for the
QGP bubbles with different Z(3) vacua. Further, for non-degenerate
vacua, even planar $Z(3)$ interfaces do not remain static, and move 
away from the region with the unique true vacuum. Thus, while for the
degenerate vacua case every closed domain wall collapses, for the
non-degenerate case this is not true any more. A closed wall enclosing
the true vacuum may expand if it is large enough so that the surface 
energy contribution does not dominate (this is essentially the same
argument as given for the bubble expansion, see Eq.(5) and the discussion
following it). 

   To see the importance of these effects we need an estimate of the
explicit symmetry breaking term arising from inclusion of quarks.
For this we use the estimates given in \cite{prsr,z3lnr}. Even 
though the estimates in ref.\cite{prsr} are given in the high temperature 
limit, we will use these for temperatures relevant for our case, i.e. 
$T \simeq 200-400$ MeV, to get some idea of the effects of 
the explicit symmetry breaking. The difference in the potential energy 
between the true vacuum with $l = 1$ and the other two vacua 
($l = e^{i2\pi/3}$, and $l = e^{i4\pi/3}$, which are degenerate with 
each other) is estimated in ref. \cite{prsr} to be,

\begin{equation}
\Delta V \sim {2 \over 3} \pi^2 T^4 {N_l \over N^3} (N^2 - 2)
\end{equation} 

 where $N_l$ is the number of massless quarks. If we
take $N_l = 2$ then $\Delta V \simeq 3 T^4$. At the bubble
nucleation temperature (which we have taken to be about $T = 200$
MeV, the difference between the false vacuum and the true vacuum
is about 150 MeV/fm$^3$ while $\Delta V$ at $T = 200$ MeV is about
four times larger, equal to 600 MeV/fm$^3$. As $T$ approaches $T_c$, 
this difference will become larger as the metastable vacuum and the 
stable vacuum become degenerate at $T_c$, while $\Delta V$ remains
non-zero. For $T$ near 250 MeV (where the barrier between the metastable 
vacuum and the stable vacuum disappears), $\Delta V$ becomes almost
comparable to the difference between the potential energy of the false 
vacuum (the confining vacuum) and the true vacuum (deconfined vacuum). 

It does not seem reasonable that at temperatures of order 200 MeV a QGP 
phase (with quarks) has higher free energy than the hadronic phase. This
situation can be avoided if the estimates of Eq.(18) are lowered
by about a factor of 5 so that these phases have lower
free energy than the confining phase. (A more desirable situation will
be when $\Delta V$ approaches zero as the confining vacuum and the 
deconfining (true) vacuum  become degenerate at $T_c$.) It is in the
spirit of the expectation that explicit breaking of Z(3) is small near 
$T_c$ for finite pion mass \cite{z3lnr}. Even with such lower 
estimates, the effects of quarks may give different nucleation probabilities
for different Z(3) vacua. However, in this paper we will ignore
this possibility. This may not be very unreasonable as for 
thick wall bubbles thermal fluctuations may be dominant in 
determining the small number of bubbles nucleated during the short 
span of time available. For very few bubbles nucleated, there may
be a good fraction of events where different Z(3) vacua may occur
in good fraction. Also note that the pre-exponential factor for
the bubble nucleation rate of Eq.(6), as given in Eq.(9) increases
with the value of $S_3(l)$. Thus, for the range of values of $S_3(l)$
for which the exponential factor in Eq.(6) is of order 1, which is
likely in our case, the nucleation rate may not decrease with larger 
values of $S_3(l)$, i.e. for
the Z(3) vacua with higher potential energies than the true vacuum.
(Of course for very large values of $S_3(l)$ the exponential term will
suppress the nucleation rate.) Thus our assumption of neglecting quark
effects for the bubble nucleation rate may not be unreasonable.

  We now consider the effect of non-zero $\Delta V$, as in Eq.(18), 
on the evolution of closed domain walls. The 
temperature range relevant for our case
is $T = 200 - 400$ MeV. In an earlier work we had numerically 
estimated the surface tension of Z(3) walls to be about 0.34 and
7.0 GeV/fm$^2$ for $T$ = 200 and 400 MeV respectively. The effects
of quarks will be significant if a closed spherical wall (with true vacuum
inside) starts expanding instead of collapsing. Again, using the
bubble free energy Eq.(5), with $\eta = \Delta V$ and $\sigma$
as the surface energy of the interface, we see that the 
critical radius $R^*$ of the  spherical wall is

\begin{equation}
R^* = {2\sigma \over \eta} \simeq 2\sigma/3T^4
\end{equation}

 For $T$ = 200 MeV and 400 MeV we get $R^* \simeq 1$ and 1.5 fm 
respectively. Though these values are not large, these are not too
small either when considering the fact that relevant sizes and times
for RHICE are of order few fm anyway. The values of $R^*$ we estimated
here  are very crude as for these sizes wall thickness is
comparable to $R^*$ hence application of Eq.(5), separating volume
and surface energy contributions, is not appropriate. Further,  
as we discussed above, the estimate in Eq.(18) which is applicable 
for high temperature limit, seems an over estimate by about
an order of magnitude at these temperatures. Thus uncertainties of 
factors of order 1 may not be unreasonable to expect. In that case
the dynamics of closed domain walls of even several fm diameter will
not be affected by the effects of quarks via Eq.(18). 

 We will see in the next section of simulation results that the domain
walls and strings typically have large velocities (e.g. about 0.5 - 0.8)
at the time of formation. These result from momentum of colliding bubble
walls and from curvature in the shape of these walls
(as well as asymmetries in the profiles of strings) at the time of 
formation. With such large velocities present, the effects of pressure 
differences between different Z(3) vacua due to quarks may become 
subdominant in studying the evolution of these structures for the short
time duration available for RHICE.
  
 With this, we will assume that for small closed walls, of order
few fm diameter, as is expected in RHICE, the quark effects in the
evolution of wall network may be neglected. 
We plan to remove these assumptions and include the
effects of non-zero $\Delta V$ due to quarks on bubble nucleation
and wall evolution in a future work.

\section{RESULTS OF THE SIMULATION}

 As we mentioned above, the number of bubbles expected to form in
RHICE is small. We first present and discuss the case of 5 bubbles which
is more realistic from the point of view of nucleation estimates
given by Eq.(6) (though a gross over-estimate for Eq.(10)). Note that a 
domain wall will form even if only two bubbles nucleate (with different 
Z(3) vacua). However, to see QGP string formation, we need nucleation 
of at least three bubbles. Next  we will discuss the case of 9 bubbles 
which is a much more optimistic estimate of the nucleation rate (even 
for Eq.(6)). Alternatively, this case can be taken as 
better representation of the case when the transition
is a cross over and bubbles only represent a means for developing a
domain structure expected after the cross-over is completed. (In this 
case only relevant energy density fluctuations, as discussed below, 
will be those arising from Z(3) walls and strings, and not the ones 
resulting from bubble wall coalescence.)

\subsection{formation and motion of extended walls}

  Fig.3-7 show the results of simulation when five bubbles are
nucleated with random choices of different Z(3) vacua inside
each bubble. Fig.3 shows a time sequence of surface plots of the 
order parameter $l(x)$ in the two dimensional lattice. Fig.3a shows
the initial profiles of the bubbles of the QGP phase embedded in
the confining vacuum with $l = 0$ at $\tau$ = 0.5 fm with the
temperature $T = $ 200 MeV. (Recall that for initial 1 fm time the 
temperature is taken to linearly increase from zero to $T_0$ = 400 
MeV.) The radial profile of each bubble is truncated with 
appropriate care of smoothness on the lattice for proper time 
evolution. Fig.3b shows the profile of each bubble at $\tau = $ 1.5
fm showing the expansion of the bubbles. Near the outer region
of a bubble the field grows more quickly towards the true vacuum.  
If bubbles expand for long time then bubble walls become 
ultra-relativistic and undergo large Lorentz contraction. 
This causes problem in simulation (see, e.g. \cite{ajit}).
In our case this situation arises at outer boundaries (for the
inner regions bubbles collide quickly). For outer regions also
it does not cause serious problem because of the use of dissipative
boundary strip (as explained above in Sect.V).

\begin{figure*}[!hpt]
\begin{center}
\leavevmode
\includegraphics[width=0.8\textwidth]{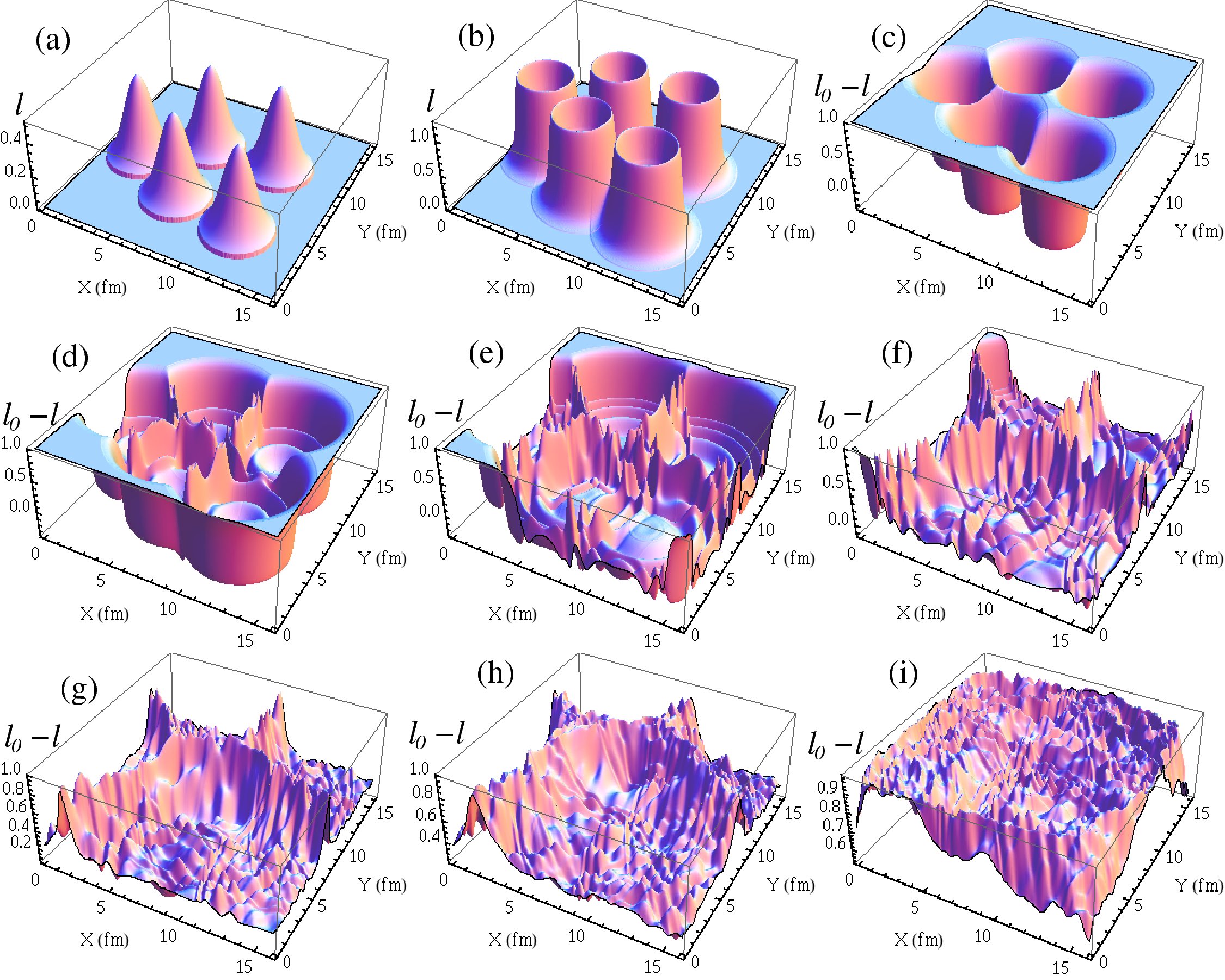} 
\end{center}
\caption{}{(a) and (b) show plots of profiles of $l$ at $\tau =$ 
0.5 fm and 1.5 fm respectively. (c)-(i) show plots of $l_0 - l$
at $\tau = $ 1.5, 2.5, 4.0, 6.0, 9.0, 11.0, and 13.7 fm. $T$
drops to below $T_c$ around at $\tau = $ 10.5 fm and $T = $ 167 MeV
at $\tau = $ 13.7 fm. Formation of domain walls and string and
antistring (at junctions of three walls) can be seen in the plots
in (e) - (h).}
\label{Fig.3}
\end{figure*}

  Figs. 3c-3i show plots of $l_0 - l$  clearly showing formation
of domain walls and strings (junctions of three walls). Here $l_0$ is 
a reference vacuum expectation value of $l$ calculated 
at the maximum temperature $T = T_0 =$ 400 MeV. Formation of domain 
walls, extending through the entire QGP region, 
is directly visible from Fig.3e
(at $\tau = 4$ fm) onwards. The temperature drops to below $Tc 
\simeq 182$ MeV at $\tau \simeq 10.5$ fm. The last plot in Fig.3i
is at $\tau = 13.7$ fm when the temperature $T = $ 167 MeV, clearly
showing that the domain walls have decayed away in the confined phase
and  the field is fluctuating about $l = 0$. 

\begin{figure*}[!hpt]
\begin{center}
\leavevmode
\includegraphics[width=0.8\textwidth]{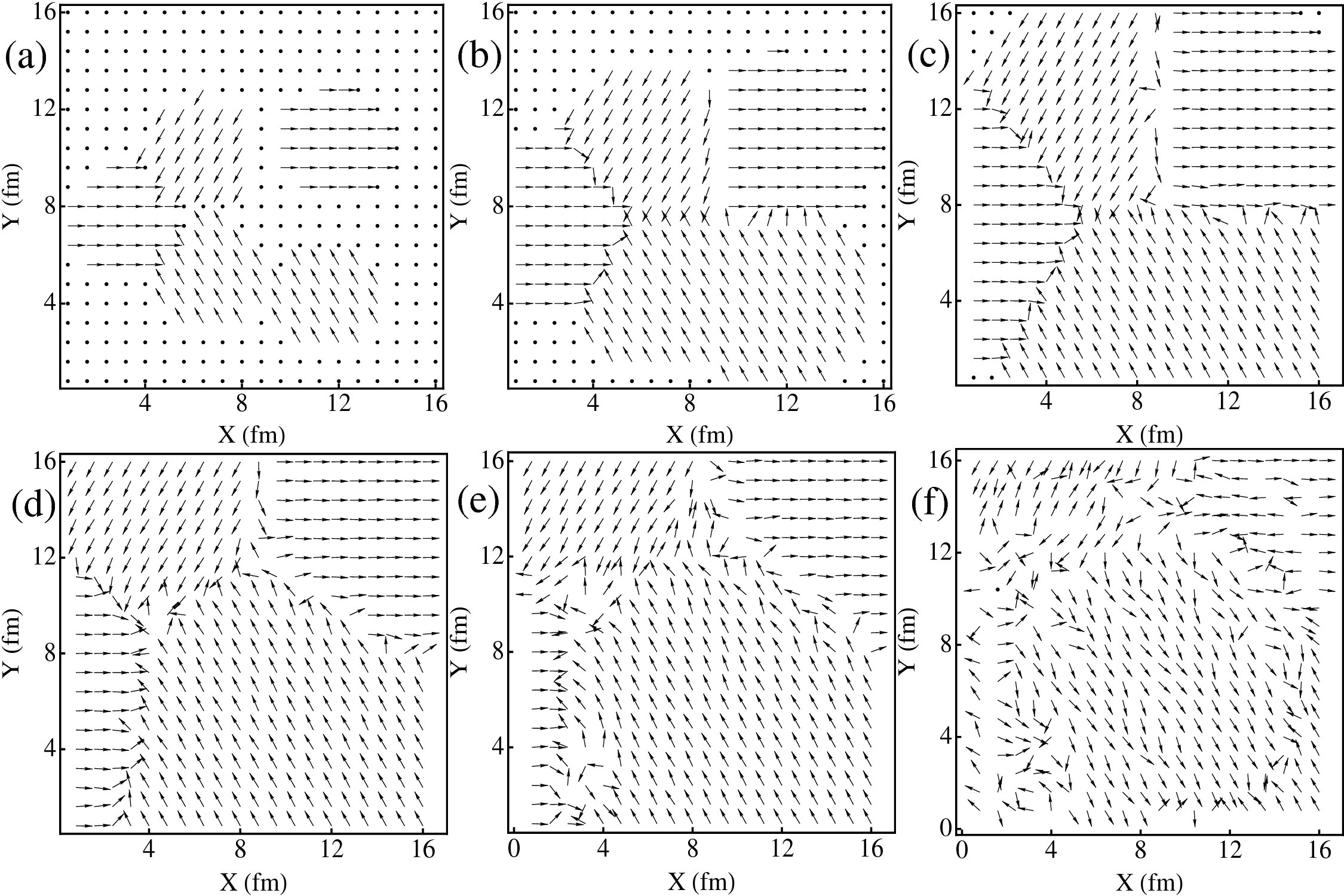} 
\end{center}
\caption{}{Plots of the phase $\theta$ of the order parameter $l$.
(a) shows the initial distribution of $\theta$ in the bubbles at $\tau =$
0.5 fm. (b) - (f) show plots of $\theta$ at $\tau = $ 2.0, 4.6,
11.0, 12.2, and 13.7 fm respectively. Location of domain walls
and the string (with positive winding) and antistring (with
negative winding) are clearly seen in the plots in (b)-(e). The
motion of the antistring and associated walls can be directly 
seen from these plots and an estimate of the velocity can be obtained.}
\label{Fig.4}
\end{figure*}

In Fig.3f-3h we see two junctions of three domain walls where the QGP 
strings form. This is seen more clearly in Fig.4 where the phase
$\theta$ of $l$ is plotted (with the convention that $\theta$ is the 
angle of the arrow from the positive $X$ axis). The domain walls
are identified as the boundaries where two different values of
$\theta$ meet, and strings correspond to the non-trivial winding of
$\theta$ at the junctions of three walls. From Fig.4b,c we clearly see
that at one of the junctions we have a string (at $X \simeq$ 5 fm, $Y
\simeq$ 8 fm) with positive winding and we have an anti-string
at $X \simeq$ 9 fm, $Y \simeq $ 8 fm with negative winding. Note
the rapid motion of the walls forming the antistring towards
positive $Y$ axis from Fig.4c (at $\tau =$ 4.6 fm) to Fig.4e
(at $\tau =$ 12.2 fm). The average speed of the antistring 
(and wall associated with that) can be directly estimated from these
figures to be about 0.5 (in natural units with $c = 1$). This
result is important in view of the discussion in the preceding
section showing that effects of pressure differences between
different Z(3) vacua, arising from quarks, may be dominated by
such random velocities present for the walls and strings at
the time of formation. The motion of the walls here is a direct
result of the straightening of the $L$ shaped wall structure
due to its surface tension. For the same reason the wall in the left
part of Fig.4b also straightens from the initial wedge shape. 
Fig.4f shows almost random variations
of $\theta$ at $\tau = 13.7$ fm when the temperature is 167 MeV,
well below the critical temperature. Though it is interesting to note
that a large region of roughly uniform values of $\theta$ still
survives at this stage.

\begin{figure*}[!hpt]
\begin{center}
\leavevmode
\includegraphics[width=0.9\textwidth]{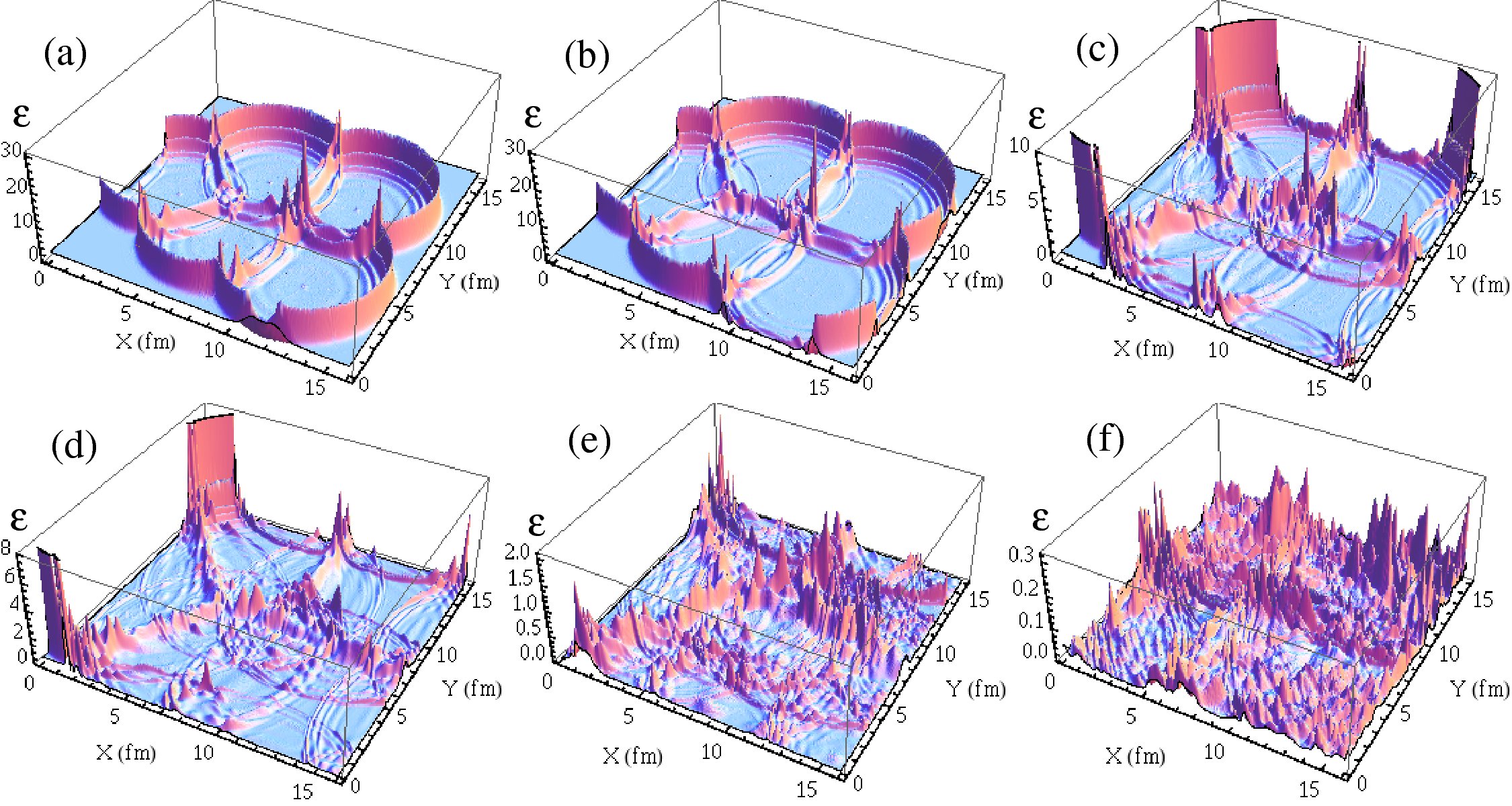} 
\end{center}
\caption{}{Surface plots of the local energy density ${\varepsilon}$
in GeV/fm$^3$. (a) - (f) show plots at $\tau = $ 3.0, 3.6, 5.0, 6.0, 
8.0, and 13.2 fm respectively. Extended domain walls can be seen from
these plots of $\varepsilon$ in (b) - (e). Small peaks in $\varepsilon$
exist at the locations of string and antistring (larger peaks
arise from oscillations of field where bubbles coalesce). Plot
in (f) is at the stage when $T = $ 169 MeV and domain walls have
decayed away.} 
\label{Fig.5}
\end{figure*}

  Fig.5 shows surface plots of the local energy density $\varepsilon$. 
$\varepsilon$ is plotted in units of GeV/fm$^3$. Although the simulation
is 2+1 dimensional representing the transverse plane of the QGP
system, we calculate energy in 3+1 dimensions by taking a thickness
of 1 fm in the central rapidity region.
Fig.5a shows plot at $\tau =$ 3 fm when bubbles have coalesced. In
Fig.5b at $\tau=$ 3.6 fm we see that bubble walls have almost decayed
(in ripples of $l$ waves) between the bubbles with same $\theta $ 
(i.e. same Z(3) vacua) as can be checked from $\theta$ plots in Fig.4.
Energy density remains well localized in the regions where domain
walls exist. Also, one can see the small peaks in the energy density where
strings and antistrings exist. Large peaks arise from oscillations of
$l$ when bubble walls coalesce, as discussed in \cite{ajit}. Large 
values of $\varepsilon$ near the boundary of the lattice are due to 
relativistically expanding bubble walls. Motion of walls and generation of
increased fluctuations in energy density are seen in Fig.5c-5e. Fig.5f
at $\tau =$ 13.2 fm (with $T$ = 169 MeV) shows that walls have decayed.
However, some extended regions of high energy density
can be seen at this stage also. 

\begin{figure*}[!hpt]
\begin{center}
\leavevmode
\includegraphics[width=0.8\textwidth]{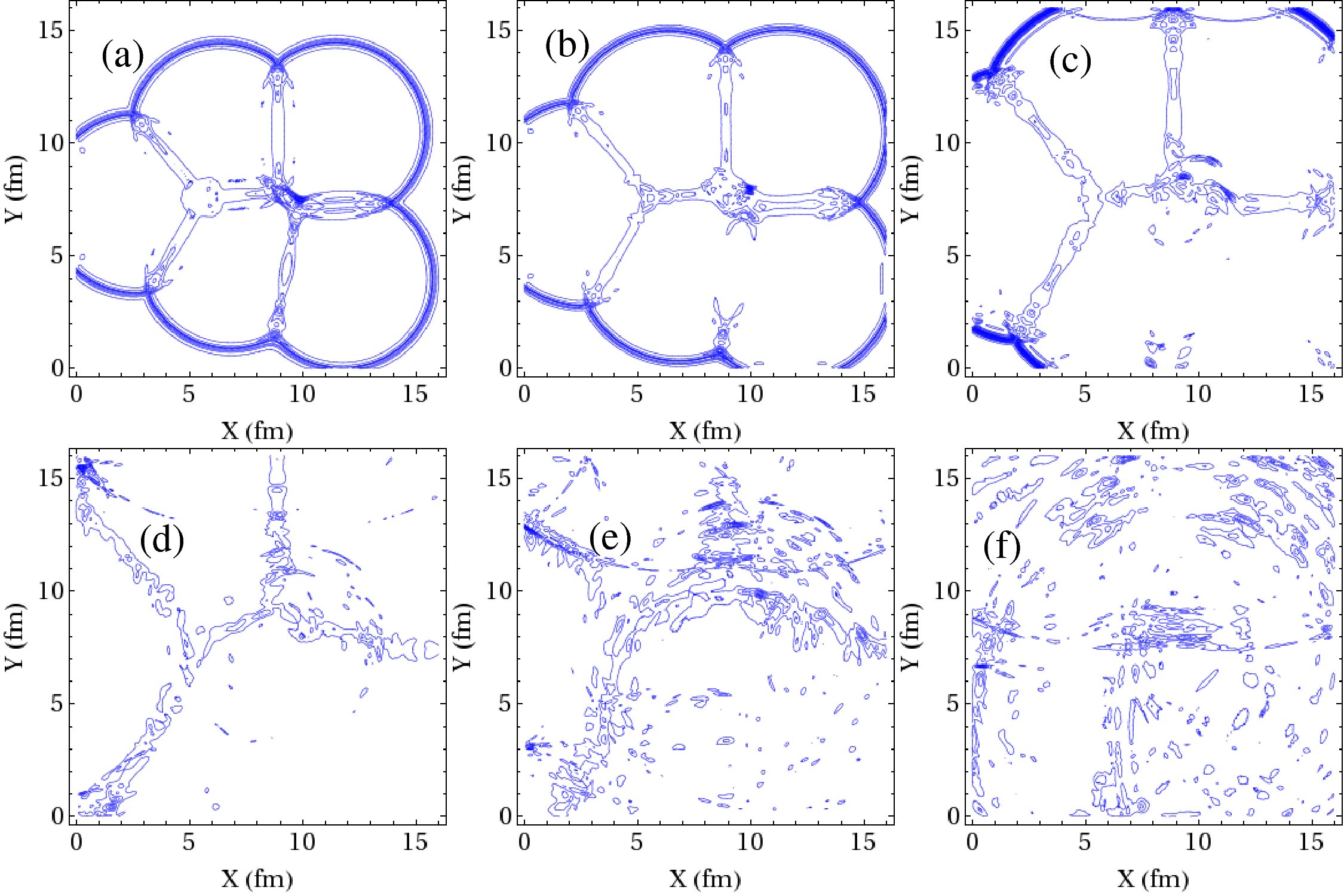} 
\end{center}
\caption{}{Contour plots of the local energy density $\varepsilon$
at different stages. Plots in (a) - (f) correspond to $\tau = $
3.0, 3.6, 5.0, 7.6, 10.2, and 13.2 fm respectively. Structure
of domains walls formed near the coalescence region of bubbles with
different $\theta$ is clear in (b),
whereas the bubble walls at lower half of $Y$ region, and near $X = $ 
10 fm, are seen to simply decay away due to same vacuum in the colliding 
bubbles.  Motion of the antistring and associated domain walls is clear
from plots in (b) - (e). The last plot in (f) is at $\tau = 13.2$ fm
when $T$ = 169 MeV.}  
\label{Fig.6}
\end{figure*}

  Fig.6 shows contour plots of energy density $\varepsilon$. Fig.6a shows
coalescence of bubbles at $\tau = $ 3 fm. Fig.6b at $\tau=$ 3.6 fm clearly
shows the difference in the wall coalescence depending on the vacua in
the colliding bubbles. Where domain walls exist, we see extended
regions of high energy density contours whereas where the two vacua in
colliding bubbles are same, there are essentially no high energy density
contours. Motion of domain walls (and strings at wall junctions) is clearly
seen in these contour plots in Fig.6b-6d. Fig.6e is at $\tau =$ 10.2 fm
when $T$ drops to $T_c$. Wall structures are still present. Fig.6f is
at $\tau = $ 13.2 fm ($T$ = 169 MeV) when walls have decayed away, though
some extended structures in contours still survive.     

  We have also calculated the variance of energy density $\Delta \varepsilon$ 
at each time stage to study how energy fluctuations change during the 
evolution. In Fig.7 we show the plot of $\Delta \varepsilon/\varepsilon$ as 
a function of proper time. Here $\varepsilon$ is the average value of energy
density at that time stage. The energy density $\varepsilon$ decreases due to 
longitudinal expansion, hence we plot this ratio to get an idea of 
relative importance of energy density fluctuations. Fig.7 shows initial
rapid drop in $\Delta \varepsilon/\varepsilon$ due to large increase in 
$\varepsilon$ during the heating stage upto $\tau$ =  1 fm, followed by 
a rise due to increased energy density fluctuations during the stage 
when bubbles coalesce and bubble walls decay, as expected. Interesting 
thing to note is a slight peak in the plot near $\tau =$ 10.5 fm when 
$T$ drops below $T_c$. This should correspond to the decay of domain 
walls and may provide a signal for the formation and subsequent decay of
such objects in RHICE.

\begin{figure*}[!hpt]
\begin{center}
\leavevmode
\includegraphics[width=0.5\textwidth]{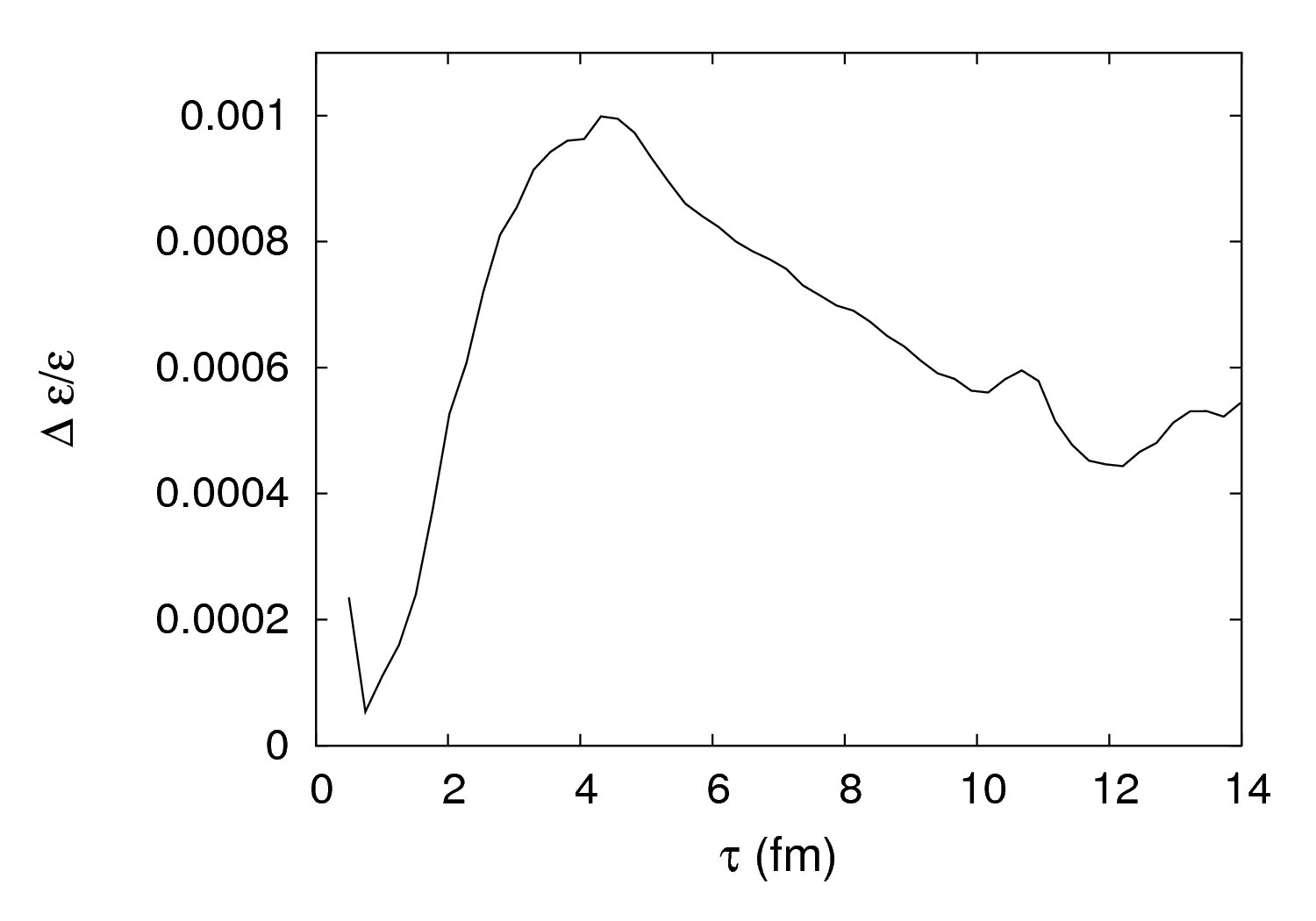} 
\end{center}
\caption{}{Plot of the ratio of variance of energy density $\Delta 
\varepsilon$ and the average energy density $\varepsilon$ as a function of 
proper time. Energy fluctuations increase during the initial stages when 
bubbles coalesce and bubble walls decay. After that there is a slow decrease
in energy fluctuation until the stage when the temperature drops below
$T_c$ and $\tau \simeq 10.5$ fm. Energy fluctuations increase after this
stage. Note small peak near the transition stage.}
\label{Fig.7}
\end{figure*}

\subsection{Formation and collapse of a closed wall}
 
  Fig.8-12 show the results of simulation where nine bubbles are
nucleated. Fig.8 shows a time sequence of surface plots of 
$l(x)$ (similar to Fig.3). Fig.8a shows the initial profiles of the 
QGP bubbles at $\tau$ = 0.5 fm with the nucleation temperature of
$T = $ 200 MeV. Fig.8b shows the profile of $l$ for the bubbles at 
$\tau = $ 1.5 fm showing the expansion of the bubbles. Fig.8c-8i
show plots of $l_0 - l$ at different stages. Noteworthy here is the
formation of a closed domain wall near the central region which is
clearly first seen in Fig.8e at $\tau = $ 5 fm. The collapse of this 
closed domain wall is seen in the subsequent plots with the closed wall
completely collapsing away in Fig.8h at $\tau = $ 9.6 fm. Only
surviving structure is an extended domain wall along the $X$ axis.
Fig.8i is at $\tau = 13.2$ fm when $T = $ 169 MeV. The domain walls
have decayed away and $l$ fluctuates about the value zero as appropriate
for the confined phase.  

\begin{figure*}[!hpt]
\begin{center}
\leavevmode
\includegraphics[width=0.8\textwidth]{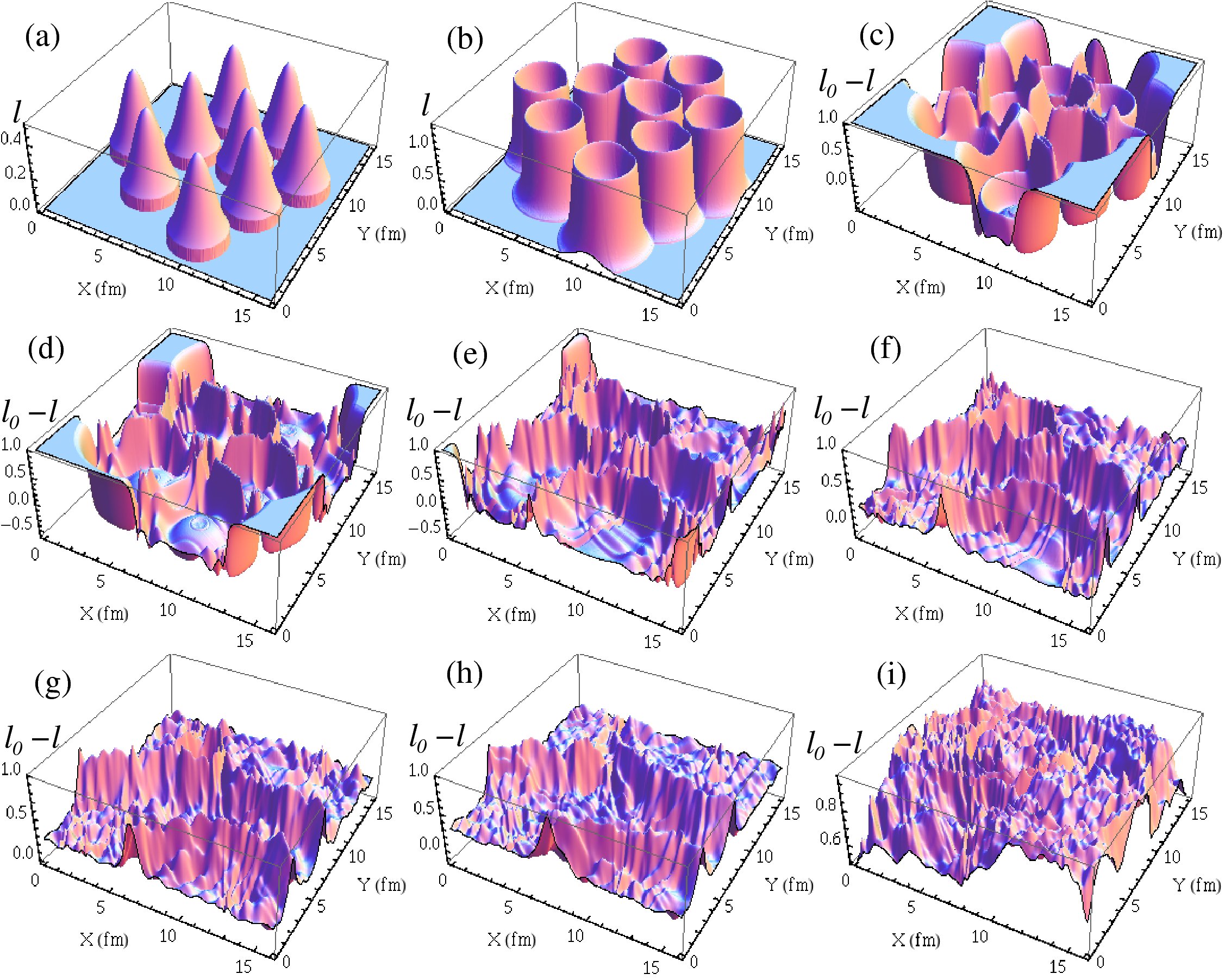} 
\end{center}
\caption{}{(a) and (b) show plots of profiles of $l$ at $\tau =$ 
0.5 fm and 1.5 fm respectively for the case when 9 bubbles are nucleated. 
(c)-(i) show plots of $l_0 - l$ at $\tau = $ 2.0, 3.0, 5.0, 7.0, 8.6, 
9.6, and 13.2 fm respectively. 
Formation of a closed domain wall is first clearly seen in the plot in (e).
This closed domain wall collapses as seen in plots in (e) through (h).
Only surviving domain wall  is an extended wall along $X$ axis in (h).
Plot in (i) is when the temperature $T = $ 169 MeV.}
\label{Fig.8}
\end{figure*}

 Fig.9 shows plots of the phase $\theta$ of $l$ at different stages.
Initial phase distribution in different bubbles is shown in Fig.9a
at $\tau = 0.5$ fm. Fig.9b shows the formation of closed, elliptical
shaped domain wall at $\tau = 2.6$ fm.  Strings and antistrings can
also be identified by checking the windings of $\theta$. The closed
domain wall collapses, and in the process becomes more circular,
as shown in the plots in Fig.9b-9h. Fig.9h shows the plot at  
$\tau = $ 9.6 fm when the closed domain wall completely collapses away,
leaving only an extended domain wall running along $X$ axis between
$Y \simeq 4 - 8$ fm. The final Fig.9i at $\tau = 13.8$ fm is when the
temperature $T =$ 167 MeV showing random fluctuations
of $\theta$ when domain walls have decayed away in the confining phase.  

\begin{figure*}[!hpt]
\begin{center}
\leavevmode
\includegraphics[width=0.8\textwidth]{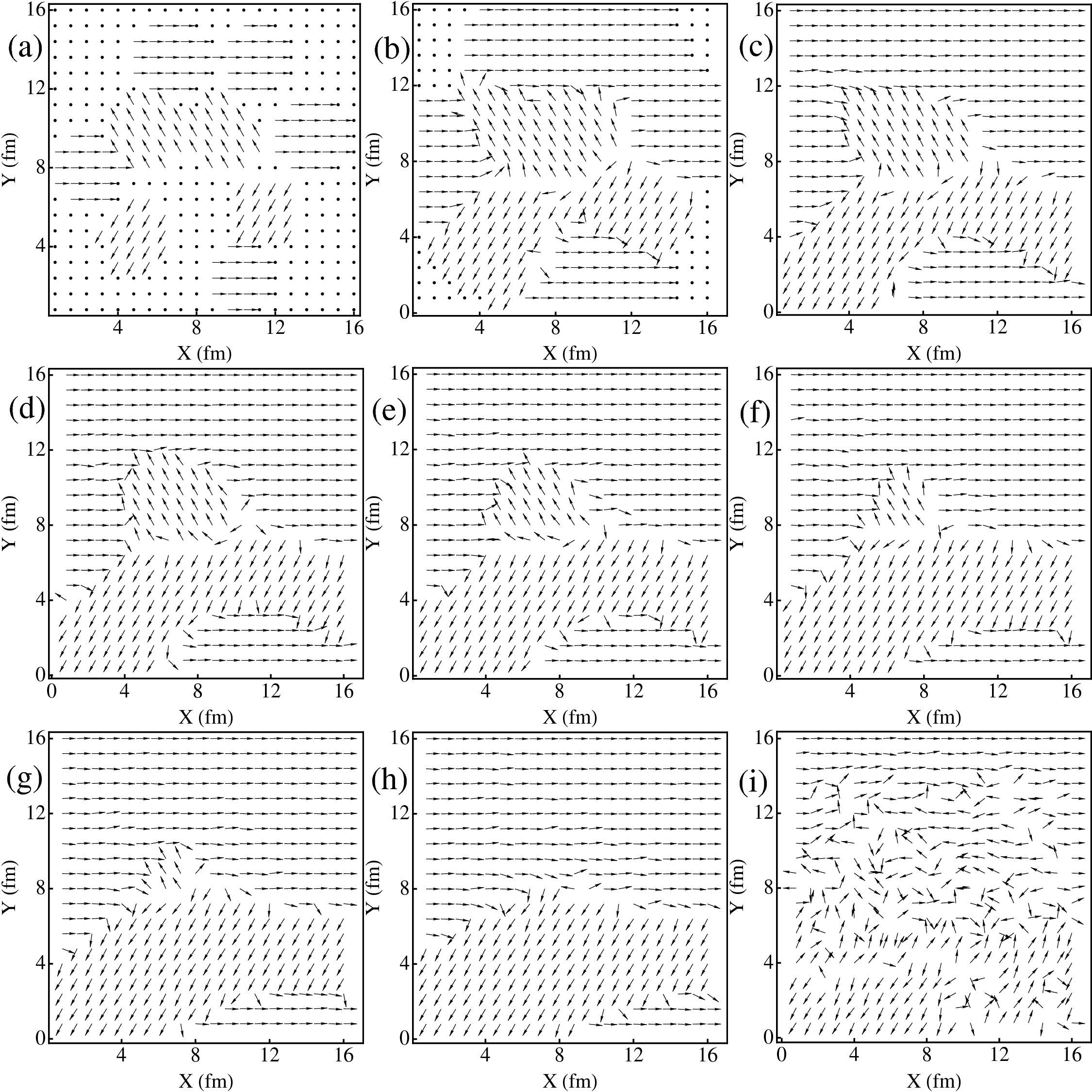} 
\end{center}
\caption{}{Plots of the phase $\theta$ of $l$.
(a) shows initial distribution of $\theta$ in the bubbles at $\tau =$
0.5 fm. (b) - (i) show plots of $\theta$ at $\tau = $ 2.6, 5.0, 6.0,
7.0, 8.0, 8.6, 9.6, and 13.8 fm respectively. (b) shows formation
of elliptical shaped closed domain wall which subsequently becomes
more circular as it collapses away by $\tau = $ 9.6 fm as shown
by the plot in (h). The plot in (i) is at $T = 167$ MeV showing 
random fluctuations of $\theta$.}
\label{Fig.9}
\end{figure*}

 Fig.10 shows the surface plot of energy density $\varepsilon$ at different
stages. Extended thin regions of large values of $\varepsilon$ are clearly 
seen in the plots corresponding to domain walls. Collapse of the closed 
domain wall is also clearly seen in Fig.10c-10g. Important thing to note
here is the surviving peak in the energy density plot at the location
of domain wall collapse. This peak survives even at the stage shown in
Fig.10i at $\tau = 13.2$ fm when $T = 169$ MeV, well below the
transition temperature. Such {\it hot spots} may be the clearest
signals of formation and collapse of Z(3) walls.   

\begin{figure*}[!hpt]
\begin{center}
\leavevmode
\includegraphics[width=0.8\textwidth]{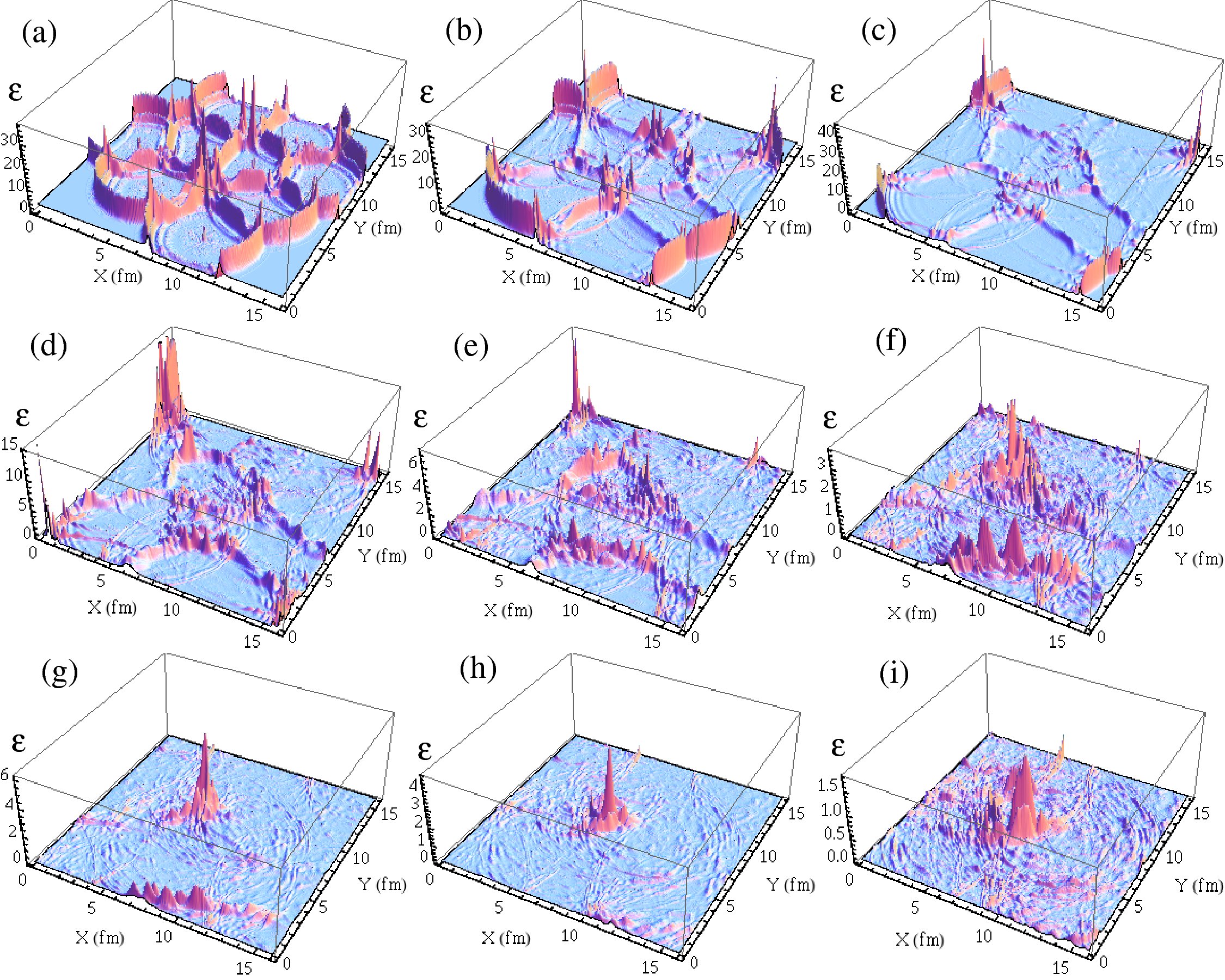} 
\end{center}
\caption{}{Surface plots of the local energy density $\varepsilon$ in 
GeV/fm$^3$.  (a) - (i) show plots at $\tau = $ 2.6, 3.5, 4.6, 5.6, 7.0, 8.6,
9.6, 11.2, and 13.2 fm respectively. Formation and subsequent collapse of
closed domain wall is clearly seen in plots in (c) through (g). Note
that the strong peak in $\varepsilon$ resulting from domain wall collapse
(the {\it hot spot}) survives in (i) when $T$ = 169 MeV.}  
\label{Fig.10}
\end{figure*}

 Contour plots of $\varepsilon$ are shown in Fig.11. Though closed domain
wall can be seen already in Fig.11b (at $\tau = $ 3.5 fm), the domain
wall is still attached to outward expanding bubble walls near 
$X = $ 3 fm, $Y = $ 12 fm which affects the evolution/motion of
that portion of the domain wall. Formation of distinct
closed wall structure is first visible in Fig.11c at $\tau =$4.6 fm. 
Subsequent plots clearly show how the domain wall becomes circular and
finally collapses away by Fig.11k at $\tau = 9.6$ fm. Note the survival
of the {\it hot spot} even at the stage shown in Fig.11{\it l} at 
$\tau = 13.2$ fm when $T = $ 169 MeV.

  One can make a rough estimate of the velocity of the closed wall
during its collapse from these plots. In Fig.11c, at $\tau = $ 4.6 fm, 
the $X$ extent of the closed wall is about 8 fm and the $Y$ extent is 
about 5 fm. The wall collapses away by the stage in Fig.11k at $\tau = $
9.6 fm. This gives rough velocity of collapse in $X$ direction to be about 
0.8 while the velocity in $Y$ direction is about 0.5. Note that here,
as well as in Fig.4 for the five bubble case, the estimate of the
wall velocity is not affected by the extra dissipation which is introduced 
only in a very thin strip (consisting of ten lattice points) near the 
lattice boundary.  Formation and collapse of such closed domain walls 
is important as the resulting hot spot can lead to important 
experimental signatures. Further, such closed domain wall structures 
are crucial in the studies of $P_T$ enhancement, especially for
heavy flavor hadrons as discussed in \cite{znb,apm} 

\begin{figure*}[!hpt]
\begin{center}
\leavevmode
\includegraphics[width=0.8\textwidth]{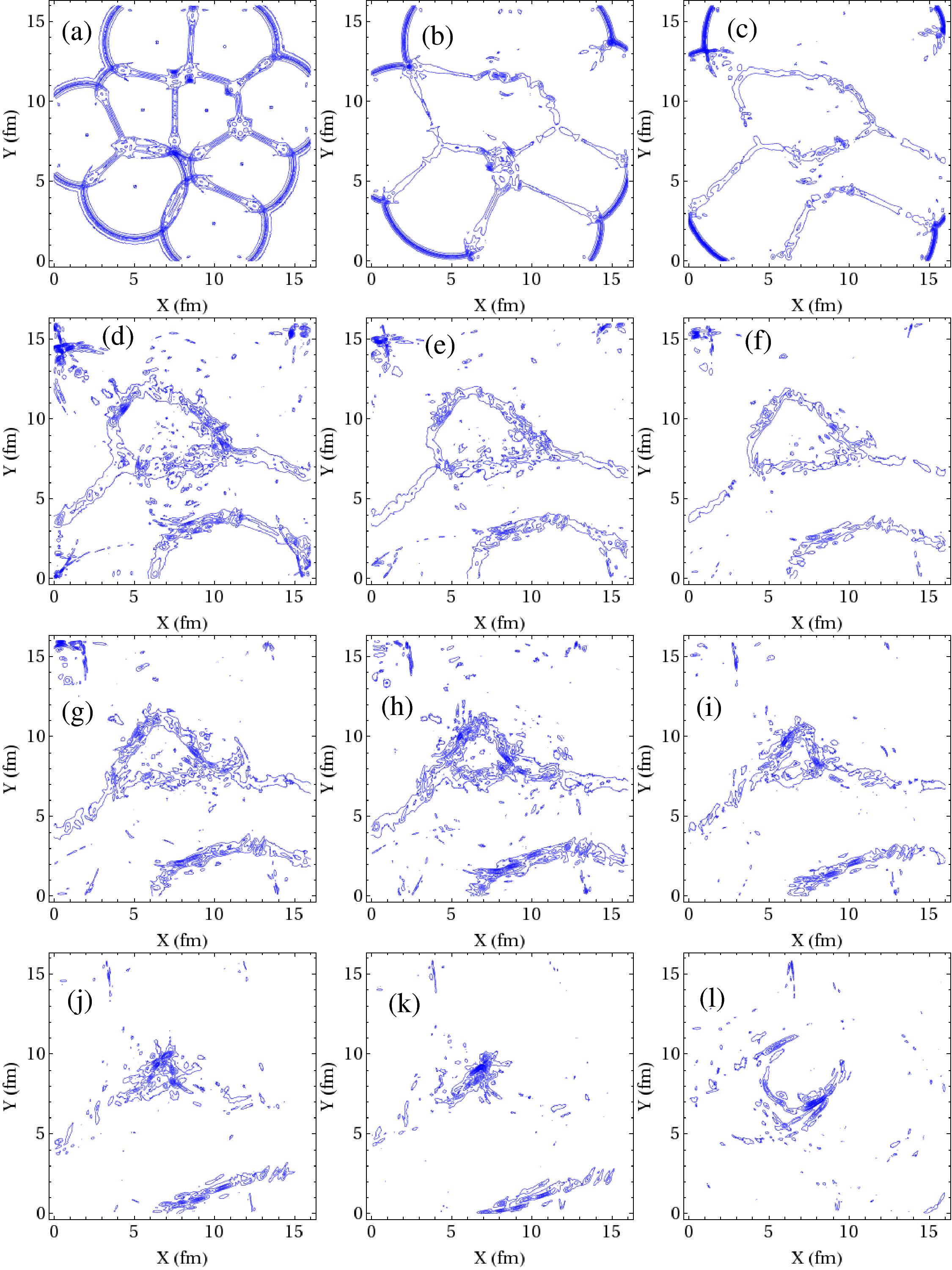} 
\end{center}
\caption{}{Contour plots of the local energy density $\varepsilon$
at different stages. Plots in (a) - (l) correspond to $\tau = $
2.6, 3.5, 4.6, 6.0, 6.6, 7.0, 7.6, 8.0, 8.6, 9.0, 9.6, and 13.2
fm respectively. Formation of distinct closed wall structure is 
first visible in (c) at $\tau =$4.6 fm. Subsequent plots show 
the collapse of this domain wall as it becomes more circular. The
wall finally collapses away in (k) at $\tau = 9.6$ fm. Note that
concentration of energy density at the location of domain wall
collapse (the {\it hot spot}) survives even at the stage shown in (l) 
at $\tau = 13.2$ fm when $T = $ 169 MeV.}
\label{Fig.11}
\end{figure*}

 Fig.12 shows the evolution of the ratio of the variance of energy density
and the average energy density. As for the five bubble case, initial
drop and rise are due to heating stage upto $\tau = $ 1 fm and
subsequent bubble coalescence and decay of bubble walls. In this case
the ratio remain roughly constant upto $\tau \simeq$ 10.5 fm which
is the transition stage to the confining phase. This is the stage
when the surviving extended domain wall starts decaying. This is also
the stage soon after the closed domain wall collapses away. The 
prominent peak at this stage should be a combined result of both
of these effects.
The large increase in the variance of energy density at this stage
should be detectable from the analysis of particle distributions
and should be a clear signal of hot spots resulting from
collapse of closed walls  and the decay of any surviving domain
walls.
 
\begin{figure*}[!hpt]
\begin{center}
\leavevmode
\includegraphics[width=0.5\textwidth]{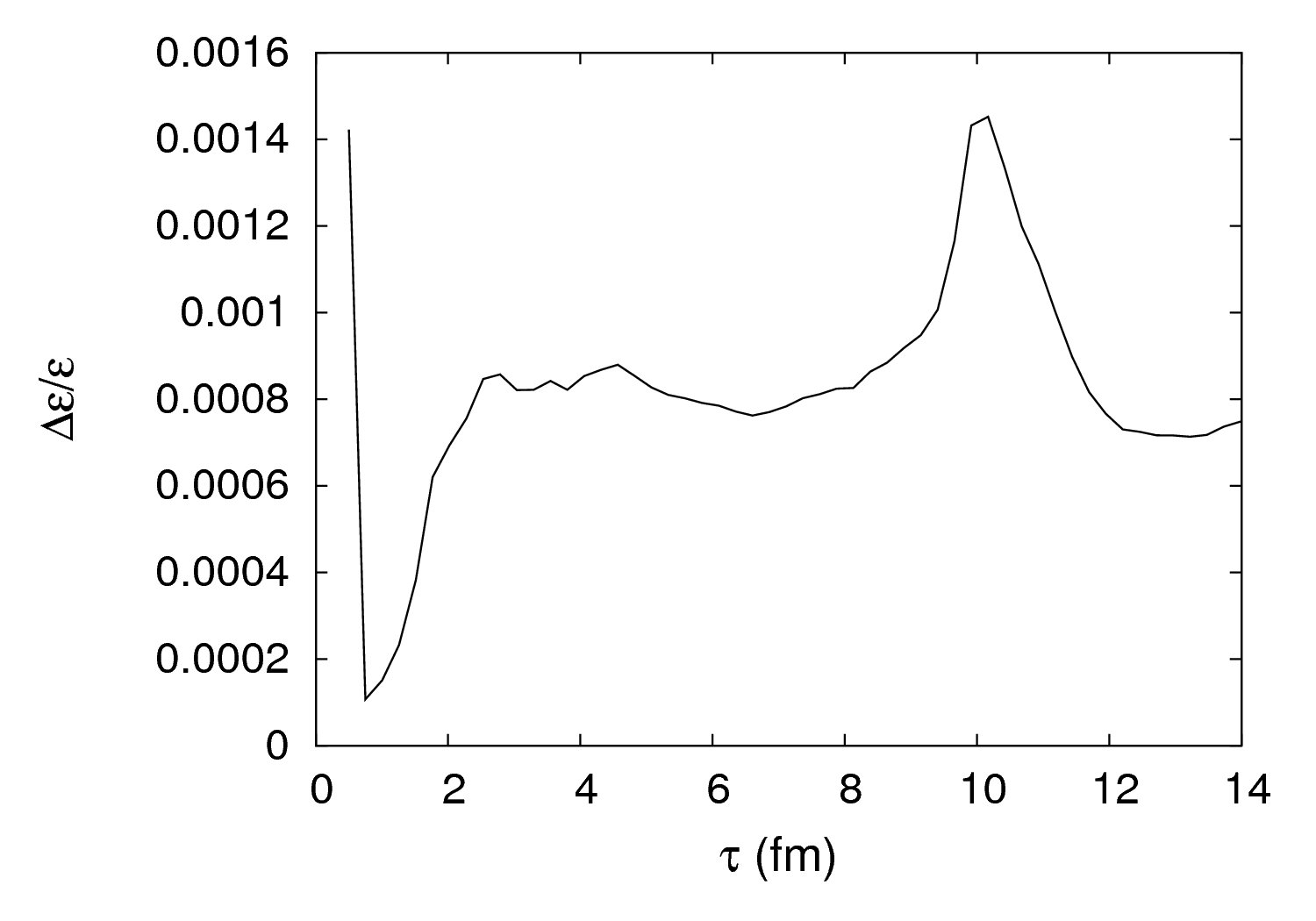} 
\end{center}
\caption{}{Plot of the ratio of variance of energy density $\Delta \varepsilon$
and the average energy density $\varepsilon$ as a function of proper time.
Energy fluctuations increase during the initial stages when bubbles
coalesce and bubble walls decay. After that $\Delta \varepsilon/\varepsilon$
remains roughly constant until the stage when the temperature drops below
$T_c$ at $\tau \simeq 10.5$ fm. This is also the stage just after the
collapse of the closed domain wall. Energy fluctuations sharply increase 
around this stage. Note the prominent peak at this stage.}
\label{Fig.12}
\end{figure*}

\section{Possible experimental signatures of Z(3) walls and strings}

 The $Z(3)$ wall network and associated strings form during the early 
confinement-deconfinement phase transition. They undergo evolution in 
an expanding plasma with decreasing temperature, and  eventually melt
away when the temperature drops below the deconfinement-confinement
phase transition temperature. They may leave their signatures in
the distribution of final particles due to large concentration of
energy density in extended regions as well as due to non-trivial
scatterings of quarks and antiquarks with these objects. 

  First, we focus on the extended regions of high energy density 
resulting from the domain walls and strings. This is clearly seen in
our simulations and some extended structures/hot spots also survive
after the temperature drops below the transition temperature $T_c$.
Note that even the hot spot resulting from the collapse of closed
domain wall in Fig.9,10 will be stretched in the longitudinal direction
into an extended linear structure (resulting from the collapse of
a cylindrical wall). We know that at RHIC energies, the final freezeout 
temperature is not too far below the transition temperature $T_c$. This 
means that the energy density concentrated in any extended (sheet like 
for domain walls and line like for strings/hot spots) regions may not 
be able to defuse away effectively. Assuming local energy density to 
directly result in multiplicity of particles coming from that region, an 
analysis of particle distribution in $P_T$ and in rapidity should be able 
to reflect any such extended regions. In this context, it will be interesting
to investigate if the ridge phenomenon seen at RHIC \cite{ridge} could
be a manifestation of an underlying Z(3) domain wall/string structure. 
Correlation of particle production over large range of rapidity
will naturally result from longitudinally extended regions of
high energy density ({\it hot spots} in the transverse plane).
Combined with flow effects it may lead to ridge like structures
\cite{ridge2,ridge}. If extended domain wall structure
survives in the transverse plane also, this will then extend to
sheet like regions in the longitudinal direction. Decay of
such a region of high energy density may directly lead to
a ridge like structure, without requiring flow effects.

  We expect non-trivial signatures resulting from the consideration of
interactions of quarks and antiquarks with domain walls. It was shown
in an earlier work \cite{znb} using generic arguments that quarks and
antiquarks should have non-zero reflection coefficients when traversing
across these domain walls. A collapsing domain wall will then concentrate
any excess baryon number enclosed, leading to formation of baryon 
rich regions. This is just like Witten's scenario for the early 
universe \cite{wtn} (which was applied for the case of RHICE in 
ref. \cite{srfc}). However, for these works it was crucial that the
quark-hadron transition be of (strong)  first order. As we have emphasized
above, in our case formation of Z(3) walls and strings will be a 
generic feature of any C-D phase transition.
Even though we have implemented it in the context of a first order
transition via bubble nucleation, these objects will form even
if the transition is a cross-over. Thus, concentration of baryons in
small regions should be expected to occur in RHICE which should
manifest in baryon concentration in small regions of rapidity
and $P_T$.

  Another important aspect of quark/antiquark reflection is that
inside a collapsing wall, each reflection increases the momentum
of the enclosed particle. When closed domain walls collapse then 
enclosed quarks/antiquarks may undergo
multiple reflections before finally getting out. This leads to a
specific pattern of $P_T$ enhancement of quarks with heavy flavors
showing more prominent effects \cite{apm}. The modification of $P_T$
spectrum of resulting hadrons can be calculated, and the enhancement 
of heavy flavor hadrons at high $P_T$ can be analyzed for the signal 
for the formation of Z(3) domain walls in these experiments \cite{apm}.
In our simulations extended domain walls also form which show
bulk motion with velocities of order 0.5. Quarks/antiquarks reflected
from such moving extended walls will lead to anisotropic momentum
distribution of emitted particles which may also provide signature of
such walls. For collapsing closed domain walls, spherical domain walls
were used for estimates  in ref. \cite{znb} and in ref. \cite{apm}.
Our simulation in the present
work provides a more realistic distribution of shapes and sizes
for the resulting domain wall network. We have estimated the velocity
of moving domain walls to range from 0.5 to 0.8 for the situations
studied. These velocities are large enough to have important
effect on the momentum of quarks/antiquarks undergoing
reflection from these walls. One needs to combine the analysis 
of \cite{znb,apm} with the present simulation to get a
concrete signature for baryon concentration and heavy flavor 
hadron $P_T$ spectrum modification. We plan to carry this out in
a future work. We also plan to study effects of spontaneous violation
of CP due to formation of these Z(3) walls in RHICE.

 Our results show interesting pattern of the evolution of  the 
fluctuations in the energy density. As seen in Fig.7 and Fig.12, 
energy density fluctuations show rapid changes during stages of 
bubble wall coalescence and during collapse/decay of domain walls. 
Even string-antistring annihilations should be contributing to 
these fluctuations. Fluctuations near the transition stage
may leave direct imprints on particle distributions. It is intriguing
to think whether dileptons or direct photons may be sensitive
to these fluctuations, which could then give a time history of
evolution of such energy density fluctuations during the early
stages as well. Even the presence of domain walls and strings
during early stages may affect quark-antiquark distributions
in those regions which may leave imprints on dileptons/direct 
photons. An important point to note is that in our model, we expect energy 
density fluctuations  in event averages (representing high energy
density regions of domain walls/strings as discussed above), as well as 
event-by-event fluctuations. These will result due to fluctuation
in the number/geometry of domain walls/strings from one event to the other
resulting from different distribution of (randomly occurring) Z(3) vacua 
in the QGP bubbles. Even the number of QGP bubbles, governed by
the nucleation probability, will vary from one event to the other 
contributing to these event-by-event fluctuations.

\section{Conclusions}

  We have carried out numerical simulation of formation of $Z(3)$ 
interfaces and associated strings at the initial confinement-deconfinement 
phase transition during the pre-equilibrium stage in relativistic 
heavy-ion collision experiments. A simple model of quasi-equilibrium 
system was assumed for this stage with an effective temperature which 
first rises (with rapid particle production) to a maximum temperature 
$T_0 > T_c$,  and then decreases due to continued plasma expansion. 

 Using the effective potential for the Polyakov loop expectation
value $l(x)$ from ref. \cite{psrsk,psrsk2} we study the dynamics
of the (C-D) phase transition in the temperature/time range when the
first order transition of this model proceeds via bubble nucleation.
As we have emphasized above, though our study is in the context
of a first order transition, its results are expected to be valid even
when the transition is a cross-over. (Though for non-zero chemical
potential the transition may indeed be of first order). The generic
nature of our results arises due to the fact that the formation of
Z(3) domain walls and  associated strings happens due to the
a general domain structure resulting after any transition (occurring
in a finite time). This is the essential physics of the Kibble 
mechanism underlying the formation of topological defects in
symmetry breaking transitions. 

 The $Z(3)$ wall network and associated strings formed during this 
early C-D transition are evolved using field equations in a plasma
which is longitudinally expanding, with decreasing temperature. We
have neglected here the transverse expansion which is a good 
approximation for the early stages near the formation stage of these
objects, but may not be a good approximation for the later parts
of simulations when temperature drops below $T_c$ and Z(3) domain walls
and strings melt away. We have studied size/shape of resulting closed 
domain wall as well as extended domain walls and have estimated the
velocities of walls to range from 0.5 to 0.8. We also calculate
the energy density fluctuations expected due to formation of these 
objects. Various experimental signals which can indicate the formation 
of these topologically non-trivial objects in RHICE  have been discussed. 
For example, existence of these objects will result in specific patterns
of energy density fluctuations which may leave direct imprints on 
particle distributions. In our model, we expect energy 
density fluctuations  in event averages (representing high energy
density regions of domain walls/strings), as well as event-by-event
fluctuations as the number/geometry of domain walls/strings and even
the number of QGP bubbles, varies from one event to the other.
Extended regions of large energy densities arising from Z(3) walls
and associated strings may be manifested 
in space-time reconstruction of hadron density (using hydrodynamic
model). The correlation of particle production over large range 
of rapidity will be expected from such extended regions. This,
combined with the flow effects (for string like regions), or
possibly directly (for sheet like extended region) may 
provide an explanation for the ridge phenomena observed at
RHIC \cite{ridge}. Also, from the reflection of quarks and antiquarks 
from collapsing domain walls, baryon number enhancement in localized
regions (due to concentration of net baryon number) as well as 
enhancement of heavy flavor hadrons at high $P_T$ is expected.

 We emphasize again that the presence of $Z(3)$ walls and string may not 
only provide qualitatively new signatures for the QGP phase in these 
experiments, it may provide the first (and may be the only possible) 
laboratory study of such topological objects in a relativistic quantum
field theory system.

\acknowledgments

  We are very grateful to Sanatan Digal, Anjishnu Sarkar, Ananta P. Mishra,
P.S. Saumia, and Abhishek Atreya for very useful comments and suggestions.
USG, AMS, and VKT acknowledge the support of the Department of Atomic 
Energy- Board of Research in Nuclear Sciences (DAE-BRNS), India, under 
the research grant no 2008/37/13/BRNS. USG and VKT acknowledge support of 
the computing facility developed by the Nuclear-Particle Physics group
of Physics Department, Allahabad University under the Center of
Advanced Studies (CAS) funding of UGC, India.



\begin{thebibliography}{99}

\bibitem{heinz} U. W. Heinz, Lectures given at 2nd Latin American 
School of High-Energy Physics, San Miguel Regla, Mexico, 1-14 Jun 2003,
published in {\it San Miguel Regla 2003, High-energy physics} 165.
e-Print: hep-ph/0407360;

\bibitem{lary} L. D. McLerran and B. Svetitsky, Phys. Rev.{\bf D24}, 
450 (1981); B. Svetitsky, Phys. Rep. {\bf 132 }, 1 (1986).

\bibitem{zn} T. Bhattacharya, A. Gocksch, C. K. Altes, and R. D.
Pisarski, Nucl. Phys. {\bf B 383}, 497 (1992); J. Boorstein and 
D. Kutasov, Phys. Rev. {\bf D 51}, 7111 (1995).

\bibitem{smlg} A. V. Smilga, Ann. Phys. {\bf 234}, 1 (1994).

\bibitem{qurk1} V.M. Belyaev, Ian I. Kogan, G.W. Semenoff, and 
N. Weiss, Phys. Lett. {\bf B277}, 331 (1992). 

\bibitem{qurk2}  C.P. Korthals Altes,  hep-th/9402028  

\bibitem{psrsk} R.D. Pisarski, Phys. Rev. {\bf D62}, 111501R (2000);
{\it ibid}, hep-ph/0101168.

\bibitem{psrsk2} A. Dumitru and R.D. Pisarski, Phys. Lett. {\bf B 504},
282 (2001); Phys. Rev. {\bf D 66}, 096003 (2002); Nucl. Phys. 
{\bf A698}, 444 (2002).

\bibitem{z3lnr} A. Dumitru, D. Roder, and J. Ruppert,
Phys. Rev. {\bf D70}, 074001 (2004).

\bibitem{z3str} B. Layek, A.P. Mishra, A.M. Srivastava,
Phys. Rev. {\bf D 71}, 074015 (2005).

\bibitem{znb} B. Layek, A.P. Mishra, A.M. Srivastava, and V.K. Tiwari,
Phys. Rev. {\bf D 73}, 103514 (2006).

\bibitem{apm}  "Z(3) Interfaces, Strings, and their consequences
in  quark-gluon plasma", A.P. Mishra, Thesis, Institute of Physics, 
Bhubaneswar, 2010. 

\bibitem{kbl} T.W.B. Kibble, J. Phys. {\bf A 9}, 1387 (1976);
Phys. Rep. {\bf 67}, 183 (1980);

\bibitem{zrk} W.H. Zurek, Phys. Rep. {\bf 276}, 177 (1996).

\bibitem{dumitru} A. Bazavov, B.A. Berg, and A. Dumitru,
Phys. Rev. {\bf D78}, 034024 (2008).

\bibitem{clmn} M.B. Voloshin, I.Yu. Kobzarev, and L.B. Okun, Yad. Fiz.
{\bf 20}, 1229  (1974) [Sov. J. Nucl. Phys. {\bf 20}, 644 (1975)]; 
S. Coleman, Phys. Rev. {\bf D15}, 2929 (1977).

\bibitem{scav} O. Scavenius, A. Dumitru and J.T. Lenaghan, 
Phys. Rev. {\bf C66}, 034903 (2002).

\bibitem{latt} G. Boyd, J. Engels, F. Karsch, E. Laermann, C. Legeland,
M. Lutgemeier, and B. Petersson, Nucl. Phys. {\bf B469}, 419
(1996); M. Okamoto et al., Phys. Rev. {\bf D60}, 094510 (1999).

\bibitem{axion} S. Chang, C. Hagmann, and P. Sikivie,
Phys. Rev. {\bf D59}, 023505 (1999); M.C. Huang and P. Sikivie,
Phys. Rev. {\bf D32}, 1560 (1985).

\bibitem{vlnkn} T. Vachaspati and A. Vilenkin, Phys. Rev.{\bf D30},
2036 (1984).

\bibitem{ajit} A.M. Srivastava Phys. Rev. {\bf D45}, R3304 (1992);
Phys. Rev. {\bf D 46}, 1353 (1992); S. Chakravarty
A.M. Srivastava Nucl. Phys. {\bf B406},795 (1993).

\bibitem{linde} A.D. Linde Nucl. Phys. {\bf B216}, 421 (1983).

\bibitem{bjorken} J.D. Bjorken, Phys. Rev.{\bf  D 27}, 140 (1983).

\bibitem{langer} J. S. Langer, Ann. Phys. (N. Y.){\bf 54}, 258 (1969);
J. S. Langer and L. A. Turski, Phys. Rev. {\bf A 8}, 3230, (1973); 
L. A. Turski and J. S. Langer, Phys. Rev. {\bf A 22}, 2189 (1980)

\bibitem{csern} L. P. Csernai and J. I. Kapusta, Phys. Rev. {\bf D 46},
1379 (1992).

\bibitem{kpst} J.I. Kapusta, A.P. Vischer, and R. Venugopalan,
Phys. Rev. {\bf C 51}, 901 (1995)

\bibitem{daniel} P. Danielewicz, Phys. Lett. {\bf 146B}, 168 (1984)  

\bibitem{rndrp} T.C. Petersen and J. Randrup, Phys. Rev. {\bf C 61},
024906 (2000).

\bibitem{prsr} J. Ignatius, K. Kajantie, and K. Rummukainen,
Phys. Rev. Lett. {\bf 68}, 737 (1992); V. Dixit and M.C. Ogilvie, 
Phys. Lett. {\bf B 269}, 353 (1991).

\bibitem{ridge} J. Adams et al. [STAR Collaboration], J. Phys. {\bf G 32},
L37,(2006) [arXiv:nucl-ex/0509030]; J.Putschke, Talk at Quark Matter 2006
for STAR collaboration), Shanghai, Nov.2006; S. A. Voloshin, Phys. Lett.
{\bf B 632}, 490 (2006)[arXiv:nucl-th/0312065].

\bibitem{ridge2} A. Dumitru, F. Gelis, L. McLerran, and R. Venugopalan,
Nucl. Phys. {\bf A 810}, 91 (2008).

\bibitem{wtn} E. Witten, Phys. Rev. {\bf D 30}, 272 (1984).

\bibitem{srfc} S. Digal, A.M. Srivastava, Phys. Rev. Lett.
{\bf 80},1841(1998).                                                       

\end{thebibliography}
\end{document}